%% file: Joint_BS_optimization_revised_2.tex
\documentclass[lettersize,journal]{IEEEtran}
\usepackage{amsmath,amsfonts}
\usepackage{algorithmic}
\usepackage{algorithm}
\usepackage{array}
\usepackage{textcomp}
\usepackage{stfloats}
\usepackage{url}
\usepackage{verbatim}
\usepackage{graphicx}
\usepackage{siunitx}
\usepackage{comment}
\usepackage{multirow}
\usepackage{makecell}
\usepackage{hyperref}
\usepackage{tikz}
\usepackage[mathlines]{lineno}
\usepackage[dvipsnames]{xcolor}
\definecolor{BrickRed}{HTML}{C00000} 

\usepackage{subfigure}
\def\BibTeX{{\rm B\kern-.05em{\sc i\kern-.025em b}\kern-.08em
    T\kern-.1667em\lower.7ex\hbox{E}\kern-.125emX}}
\usepackage{balance}
\usepackage{acro}
\include{acronyms}

\usepackage{fancyhdr}
\fancypagestyle{empty}{fancy}

\begin{document}
\title{Joint BS Deployment and Power Optimization for Minimum EMF Exposure with RL in Real-World Based Urban Scenario}
\author{Xueyun Long,~\IEEEmembership{Graduate Student Member,~IEEE,} Yueheng Li,~\IEEEmembership{Member,~IEEE,} Mario Pauli,~\IEEEmembership{Senior Member,~IEEE,} Benjamin Nuss,~\IEEEmembership{Senior Member,~IEEE} and Thomas Zwick,~\IEEEmembership{Fellow,~IEEE}
\vspace{-1.7\baselineskip}}


\maketitle
\begin{abstract}Base station (BS) deployment remains a critical task with successive wireless communication generations and increasing data rates demands, while the electromagnetic field (EMF) exposure is often underrated, yielding potential health implications. Therefore, this paper proposes a workflow that adjusts BS deployment and radiated power in a 3D urban scenario to jointly consider EMF exposure and coverage. To achieve this ambition, firstly, a novel least-time shoot-and-bounce ray (SBR) ray-launching (RL) tool is developed to improve computational efficiency, and simultaneously enhance diffraction modeling for accurate EMF exposure calculation, validated with real-world measurements. To efficiently extend the computation across the target urban area, the adaptive grid refinement (AGR) algorithm is designed based on the spatial stability of the effective channel while accounting for BS beamforming, enabling global estimation of EMF exposure and signal coverage.
Subsequently, to provide a more comprehensive evaluation of EMF exposure under practical network operations, a system-level abstraction further accounts for the actual maximum transmit power, intercell interference, and channel state information imperfection on the BS side, while capturing mobility over the EMF exposure averaging interval on the user equipment side.
Upon the aforementioned aspects, the coverage-guaranteed EMF exposure minimization problem is formulated in a realistic and accurate manner, and solved by a geometry-aware algorithm that enables joint BS deployment and power optimization based on deterministic channel simulations under practical deployment constraints, completing the proposed workflow.
Simulation results illustrate the application of the workflow in the considered urban scenario, while comparison with an empirical baseline highlights that the workflow mitigates EMF exposure underestimation and yields BS deployment configurations better aligned with real-world conditions.
\end{abstract}


\section{Introduction}
\label{introduction}
Over recent decades, wireless communication technologies have advanced substantially and now constitute an indispensable part of everyday life. The proliferating number of mobile subscribers has driven demand for higher data rates to ensure satisfactory \ac{QoS} for \ac{UE} \cite{rao20215g}. To fulfill such requirement, base stations (BSs) might be gradually constructed or rebuilt, yielding BS deployment optimization an extensively studied field. Among common objectives, link quality is often pursued by maximizing \ac{SINR} via selecting optimal BS sites from candidates using heuristic method \cite{amaldi2003planning}. Coverage is maximized by jointly optimizing the number and placement of BSs with a geographic information system (GIS)-coupled artificial immune system heuristic algorithm \cite{wang2020optimizing}. Another recurring objective is reducing power consumption, achieved by solving continuous BS placement problem with pattern search \cite{khalek2011optimization} or by jointly optimizing the transmit power of the common pilot channel (CPICH), antenna tilt, and azimuth via simulated annealing \cite{siomina2006automated}. In a heterogeneous network, cost minimization is attained by selecting the cheapest combination of macro and micro BSs with a genetic algorithm \cite{valavanis2014base}. 
In particular, to further enhance the above metrics, beamforming is employed to enhance the \ac{QoS} and is expected to play a crucial role in the future \cite{chiaraviglio2021pencil}. Paradoxically, as technology advances, public concern regarding \ac{EMF} exposure continues to grow, yet it is often underrated and should be explicitly incorporated into BS deployment. To regulate \ac{EMF} exposure, the guidelines proposed by \ac{ICNIRP} impose limits \cite{international1998guidelines}, which are among the most widely adopted standards, as referenced in the 26th BImSchV in Germany \cite{des1996immissionsschutzgesetzes}. 
The guidelines specify reference levels for limiting \ac{EMF} exposure of general public, including \ac{E-field} strength, which is the benchmark for \ac{EMF} exposure in this paper. 
In any given scenario, the measured or calculated \ac{E-field} strength should be averaged over a 6-minute period and must not exceed the corresponding limit. Overall, jointly considering \ac{EMF} exposure and \ac{QoS} when deploying BS is crucial for building safety margins for real-world uncertainty, facilitating network expansion and improving public acceptance of mobile communication developments.

Seeking for the fundamental of the aforementioned BS deployment, the existing EMF exposure modeling is investigated as follows. The first idea is to model EMF exposure with empirical channel models. The extrapolation method for EMF exposure under \ac{NLoS} conditions has been studied in \cite{bechta2022analysis}. A closed-form downlink EMF exposure expressions is derived within a \ac{SG} framework in \cite{9462948}, calibrating the model parameters to measurements. In general, empirical channel models are easy to adapt using equations and aim to reproduce characteristics observed in large-scale channel measurements, yet come with limitations. One drawback is that typically only path loss is provided, lacking the space-time and angle-delay information needed for \ac{MIMO} beamforming. Another drawback is that the accuracy degrades outside the environments for which parameters were obtained \cite{yun2015ray}. Consequently, empirical channel models only provide rough estimates for \ac{EMF} exposure, insufficient to support practical BS deployment.
The second idea is to model EMF exposure with machine learning (ML) method, which is rapidly gaining interest. Recent examples include a physics-informed ensemble of decision trees \cite{10533177}, random-forest regression across multiple carrier frequencies \cite{shi2024electromagnetic}, and a reconfigurable neural network architecture (RAWA-NN) framework to estimate temperature rise and absorbed power density \cite{yao2024prediction}. However, its limited interpretability makes BS deployment solutions difficult to communicate with the public from an \ac{EMF} exposure perspective, which requires transparent and traceable reasoning. Besides, due to the limited data diversity and the ML model's poor generalization ability, its estimation may overlook signal variations caused by building heights, and violate physical constraints \cite{wang2025wireless}, thereby undermining the reliability of BS deployment and increasing the risk of exceeding \ac{EMF} exposure limits.

To address the limitations of the above mentioned methods, deterministic channel models for \ac{EMF} exposure calculation are established. 
In this manner, physical wave propagation in 3D scenarios is computed using appropriate formulas for different phenomena based on \ac{GO} to model multipath propagation, and geometrical paths are typically determined by the \ac{RL} or \ac{RT} method \cite{yun2015ray}. However, the \ac{RT} implementation that relies on the image method in \cite{fugen2006capability} requires significant computational effort as geometric complexity grows, leading to low efficiency in urban scenarios.
In addition, existing \ac{RL} tools either omit diffraction \cite{felbecker2012electromagnetic} or capture only limited diffraction paths \cite{9753133}, such that diffraction has not been thoroughly considered, which yields inaccurate \ac{EMF} exposure computations.
Moreover, to represent real-world operation, a credible \ac{EMF} exposure model should account for network behaviors, as the \ac{UE} side measurements implied that the radiated power from 5G BS is typically below the maximum value \cite{9521570}. From the network operator side, direct monitoring of 5G BS output \ac{EIRP} in live networks showed that assuming maximum power in a fixed beam over long-time intervals leads to unrealistic EMF exposure assessments \cite{10500988}. 
In summary, a high-fidelity \ac{EMF} exposure model with more efficient computation, strengthened diffraction modeling, and incorporation of realistic network behaviors is required to guide the deployment of BS.

With an \ac{EMF} exposure model in place, the BS density is optimized using a \ac{SG} model to minimize \ac{EI} for both uplink and downlink, with the radiation-sensitive areas considered \cite{10047969}. In \cite{crainic2006tackling}, the BS deployment is optimized to minimize the \ac{EMF} exposure, utilizing \ac{FSPL}, COST231-Okumura-Hata, and COST231-Walfisch-Ikegami. However, these approaches predominantly rely on empirical channel models, and are inconsistent with the expectations of the aforementioned high-fidelity \ac{EMF} exposure model. Therefore, novel algorithms should be developed to establish a complete BS deployment workflow that jointly considers \ac{EMF} exposure and \ac{QoS}.

To resolve the aforementioned open aspects, this paper presents a workflow that adjusts the BS deployment and radiated power in a real-world based urban scenario to jointly consider EMF exposure and coverage. The major contributions are as follows:
\begin{itemize}
\item{A novel \ac{RL} simulation tool, incorporating a least-time \ac{SBR} algorithm, is developed to improve computational efficiency and enhance diffraction modeling for accurate \ac{EMF} exposure calculation. Validation with real-world measurements indicates that the \ac{RL} tool generates a reliable deterministic channel model for urban scenarios.}
\item{The \ac{AGR} algorithm is designed to efficiently estimate the \ac{EMF} exposure and signal coverage across the urban area. Based on the spatial stability of the effective channel while accounting for BS beamforming, the \ac{AGR} algorithm adjusts the grid resolution accordingly, thereby reducing computational effort and preserving high estimation accuracy.}
\item{A comprehensive \ac{EMF} exposure model is presented to provide a more realistic assessment in practical communication network operation through system-level abstractions. It accounts for the actual transmit power, interference from neighboring cells considering beamforming and \ac{CSI} imperfection. On the \ac{UE} side, it captures the mobility that conforms to urban topology.}
\item{The workflow is finalized by formulating the coverage-guaranteed EMF exposure minimization problem, which is solved using a geometry-aware algorithm that enables joint BS deployment and power optimization based on deterministic channel simulations under practical deployment constraints. Simulation comparison with an empirical baseline in the considered urban scenario demonstrates that the proposed workflow mitigates the underestimation of EMF exposure and deployment discrepancies arising from mismatches between empirical channel model and geometry-dependent propagation conditions.}
\end{itemize}

The remainder of this paper is organized as follows. Section~\ref{system_model} describes ray-based system model and Section~\ref{SBR_RL_algo} introduces the least-time \ac{SBR} \ac{RL} algorithm and the real-world validation measurements. Section~\ref{agr_algo} presents the \ac{AGR} algorithm. Section~\ref{BS_place_problem_and_NM} formulates the problem and presents the proposed solution. Section~\ref{sec:experiment_results} provides the optimization results, and Section~\ref{sec:conc} concludes the paper.

\section{Ray-based System Model Formulation}
\label{system_model}
In this section, the ray-based system model in the multipath scenario is presented, including detailed calculations for different propagation phenomena and the computation of \ac{E-field} strength at the \ac{Rx} position, which serves as the \ac{EMF} exposure metric used in BS deployment optimization.

\subsection{Scenario Description}
\begin{figure*}[!t]
\centering
\includegraphics[width=0.8\textwidth, trim=1mm 3mm 1mm 8mm,clip]{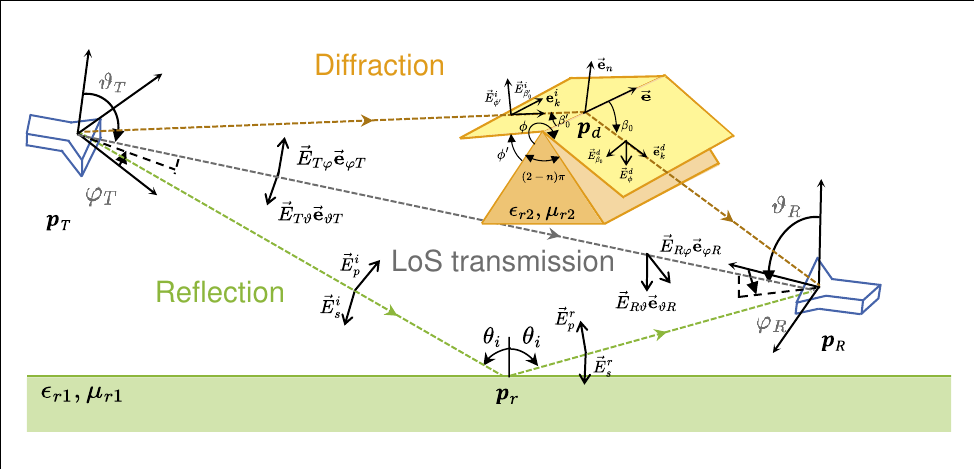}
\vspace{-1\baselineskip}
\caption{Schematic of multipath scenario for different wave propagation phenomena with coordinates and vector definition.}
\label{prop_phe}
\vspace{-1\baselineskip}
\end{figure*}

The proposed scenario, depicted by the ray-based model in Fig.~\ref{prop_phe}, includes the \ac{Tx} and \ac{Rx} at position with 3D coordinates $\boldsymbol{p}_{T} = [x_T,y_T,z_T]$ and $\boldsymbol{p}_{R} = [x_R,y_R,z_R]$, representing the simplified model of BS and \ac{UE} in a typical urban environment. Due to the large distances between BS and \ac{UE}, only the far-field propagation is considered. Geometrical rays between them in Fig.~\ref{prop_phe} include \ac{LoS} transmission path, reflection path from large smooth surfaces such as the ground or building facade, and diffraction path caused by building edges or rooftops. Based on \ac{GO}, \ac{EM} wave propagation is modeled by treating each path in Fig.~\ref{prop_phe} as a ray that carries energy and travels in a straight line through a homogeneous medium \cite{yun2015ray}. To focus exclusively on the calculations, the rays in Fig.~\ref{prop_phe} are assumed to be predetermined in this section, with the determination algorithm detailed in Section \ref{SBR_RL_algo}. 

\subsection{E-field Strength and Received Power at \ac{Rx}}
\label{e_pr_calculation}
Considering a \ac{Tx} at the origin of the local spherical coordinates in Fig.~\ref{prop_phe}, the E-field strength at a random position $(s,\vartheta_T, \varphi_T)$ in free-space is formulated according to \cite{geng2013planungsmethoden}
\begin{equation}
    \vec{E}_T(s, \vartheta_T, \varphi_T) = \sqrt{\frac{P_T G_T Z_{F0}}{2\pi}} \frac{1}{s} e^{-jk_0 s} \vec{C}_T(\vartheta_T, \varphi_T),
    \label{E_field_around_T}
\end{equation}
where $P_T$, $G_T$, $\vec{C}_T(\vartheta_T, \varphi_T)$ are the transmit power, antenna gain and complex characteristic pattern at direction $(\vartheta_T, \varphi_T)$, which is the azimuth and elevation angle in spherical coordinates. Besides, $Z_{F0}$ is the free-space impedance, $1/s$ is the attenuation factor related to distance $s$ and $e^{-jk_0 s}$ is the phase shift with wave number $k_0= 2\pi/\lambda_0$ and wavelength $\lambda_0 = c_0/f_0$, where $c_0$ is light speed and $f_0$ is the carrier frequency. Ray will then go through multipath and gathered at \ac{Rx}, and for each ray $m$, its amplitude is represented as
\begin{equation}
    V_{R,m} \!=\! \sqrt{\frac{\lambda_0^2 G_R \Re (Z_{AR})}{\pi Z_{F0}}} \, \!\vec{C}_ {R}(\vartheta_{R,m}, \!\varphi_{R,m})\!\vec{E}_{R,m}(\vartheta_{R,m}, \!\varphi_{R,m}),
    \label{U_R}
\end{equation}
where $\vec{E}_{R,m}(\vartheta_{R,m}, \varphi_{R,m})$ is the complex E-field strength at the \ac{Rx} with angle $(\vartheta_{R,m}, \varphi_{R,m})$ in local \ac{Rx} coordinates. Same as \ac{Tx}, $\vec{C}_R(\vartheta_{R,m}, \varphi_{R,m})$ is the complex characteristic pattern for \ac{Rx} at direction $(\vartheta_{R,m}, \varphi_{R,m})$, $G_R$ is the \ac{Rx} antenna gain and $\Re (Z_{AR})$ is the real part of \ac{Rx} antenna impedance. In order to derive the E-field strength $\vec{E}_R$ at the \ac{Rx}, $\vec{E}_{R,m}(\vartheta_{R,m}, \varphi_{R,m})$ must first be transformed into the global Cartesian coordinates along $(x,y,z)$ as $\vec{E}_{R,m}(x,y,z) =(\vec{E}_{Rx,m},\vec{E}_{Ry,m},\vec{E}_{Rz,m})$, and then in total $M_p$ multipath components are summed
\begin{equation}
    \vec{E}_R = \sum_{m=1}^{M_p} \vec{E}_{Rx,m} \, \vec{\boldsymbol{e}}_x + \sum_{m=1}^{M_p} \vec{E}_{Ry,m} \, \vec{\boldsymbol{e}}_y + \sum_{m=1}^{M_p} \vec{E}_{Rz,m} \, \vec{\boldsymbol{e}}_z,
    \label{E_R_vec}
\end{equation}
where $\vec{\boldsymbol{e}}_x$, $\vec{\boldsymbol{e}}_y$, $\vec{\boldsymbol{e}}_z$ are the unit vectors along $x,y,z$ directions. The effective electric field strength $E_R$ and the corresponding field strength level $E[\text{dB}\mu \text{V/m}]$ is therefore determined as \cite{geng2013planungsmethoden}
\begin{equation}
\label{E_dbuvm}
\begin{aligned}
&E_R = \frac{1}{\sqrt{2}} \sqrt{E_{Rx}^2 + E_{Ry} ^2 +  E_{Rz}^2}, \\
&E[\text{dB}\mu \text{V/m}] = 20 \log ( \frac{E_R }{1 \, \mu\text{V/m}}). 
\end{aligned}
\end{equation}
The metric used for evaluating \ac{EMF} exposure is exactly $E[\text{dB}\,\mu \text{V/m}]$ in (\ref{E_dbuvm}).
Finally, the sum of $M_p$ multipath amplitude leads to the total amplitude with $ V_{R} = \sum_{m=1}^{M_p} V_{R,m}$ and the received power $P_R$ at \ac{Rx} is calculated based on (\ref{U_R})
\begin{align} 
    P_R & = \frac{\left| \sum_{m=1}^{M_p} V_{R,m} \right|^2}{8 \Re(Z_{AR})} \nonumber \\
        &= \frac{\lambda_0^2G_R}{8 \pi Z_{F0}}\! \cdot \! \left| \sum_{m=1}^{M_p} \vec{C}_R(\vartheta_{R,m}, \varphi_{R,m}) \vec{E}_{R,m}(\vartheta_{R,m}, \varphi_{R,m}) \right|^2\!.
    \label{P_R}
\end{align}
\subsubsection{\ac{LoS} Transmission and Reflection Calculation}
\label{subsec:ref}
Assume that the \ac{LoS} path exists between \ac{Tx} and \ac{Rx} as shown in Fig.~\ref{prop_phe}, the E-field strength at \ac{Rx} position is calculated directly using (\ref{E_field_around_T}) with $s$ being the distance between $\boldsymbol{p}_{T}$ and $\boldsymbol{p}_{R}$. The received power for LoS transmission is then determined by substituting (\ref{E_field_around_T}) into (\ref{P_R}), resulting in
\begin{equation}
    P_{R,LoS} = (\frac{\lambda_0}{4 \pi})^2 G_T G_R P_T |\vec{C}_R(\vartheta_{R}, \varphi_{R}) \frac{e^{-jk_0 s}}{s} \vec{C}_T(\vartheta_T, \varphi_T)|^2.
    \label{los_pr}
\end{equation}
It should be stressed that the E-field for both \ac{Tx} and \ac{Rx} is defined in the local coordinate with unit vector $(\vec{\boldsymbol{e}}_{\vartheta T }, \vec{\boldsymbol{e}}_{\varphi T})$ and $(\vec{\boldsymbol{e}}_{\vartheta R}, \vec{\boldsymbol{e}}_{\varphi R})$. Therefore, in order to calculate (\ref{los_pr}), coordinates transition needs to be included.

When a ray intersects with a surface at a reflection point $\boldsymbol{p}_{r}$ as in Fig.~\ref{prop_phe}, the Fresnel reflection coefficients for different polarization $R_s(\theta_i)$ and $R_p(\theta_i)$ are calculated according to \cite{geng2013planungsmethoden}, using the complex relative permeability $\mu_{r1}$, permittivity $\epsilon_{r1}$ of the reflecting object, and the incident angle of the ray $\theta_i$. 
The E-field strength after $\boldsymbol{p}_{r}$ for perpendicular $\vec{E}_s^r$ and parallel $\vec{E}_p^r$ polarization is formulated as
\begin{equation}
\begin{pmatrix}
\vec{E}_s^r(\boldsymbol{p}_{r}) \\
\vec{E}_p^r(\boldsymbol{p}_{r})
\end{pmatrix}
=
\begin{bmatrix}
R_s(\theta_i) & 0 \\
0 & R_p(\theta_i)
\end{bmatrix}
\begin{pmatrix}
\vec{E}_s^i(\boldsymbol{p}_{r}) \\
\vec{E}_p^i(\boldsymbol{p}_{r})
\end{pmatrix},
\label{ref_E_1}
\end{equation}
where $\vec{E}_{s}^i(\boldsymbol{p}_{r})$ and $\vec{E}_{p}^i(\boldsymbol{p}_{r})$ are the incident E-field strength before $\boldsymbol{p}_{r}$. According to the \ac{GO} method, $\boldsymbol{p}_{r}$ is chosen to be the reference point of astigmatic beam and due to the spherical wave assumption, E-field strength at distance $s$ after $\boldsymbol{p}_{r}$ is represented as
\begin{equation}
\begin{pmatrix}
\vec{E}_s^r(s) \\
\vec{E}_p^r(s)
\end{pmatrix}
=
\frac{s'}{(s' + s)} e^{-jk_0 s}
\begin{bmatrix}
R_s(\theta_i) & 0 \\
0 & R_p(\theta_i)
\end{bmatrix}
\begin{pmatrix}
\vec{E}_s^i(\boldsymbol{p}_{r}) \\
\vec{E}_p^i(\boldsymbol{p}_{r})
\end{pmatrix}.
\label{modified_ref_E}
\end{equation}
For single reflection, $s'$ is the distance from $\boldsymbol{p}_{T}$ to $\boldsymbol{p}_{r}$. For multiple reflections, (\ref{modified_ref_E}) will be used iteratively with $s'$ being the distance from the last reflection point to the current reflection point. After projecting the polarized components $\vec{E}_s^r(s)$ and $\vec{E}_p^r(s)$ into Cartesian coordinates, the components are summed to derive $\vec{E}_{R,m}(x,y,z)$ for this reflection path.

\subsection{Diffraction Calculation} 
\label{subsec:diff}
The diffraction coefficient is determined using the heuristic \ac{UTD} method for dielectric rough wedges, as proposed by Luebbers \cite{geng2013planungsmethoden}
\begin{equation}
\begin{aligned}
D_h &= \frac{-e^{-j\frac{\pi}{4}}}{2n\sqrt{2\pi k_0}\sin\beta_0'}
\left( D_0^{\text{ISB}} + D_n^{\text{ISB}} + R_0^h D_0^{\text{RSB}} + R_n^h D_n^{\text{RSB}} \right), \\
D_s &= \frac{-e^{-j\frac{\pi}{4}}}{2n\sqrt{2\pi k_0}\sin\beta_0'}
\left( D_0^{\text{ISB}} + D_n^{\text{ISB}} + R_0^s D_0^{\text{RSB}} + R_n^s D_n^{\text{RSB}} \right).
\end{aligned}
\label{diffraction_coefficient}
\end{equation}
Assuming the diffraction path exists in Fig.~\ref{prop_phe}, the diffraction coefficient is represented using soft ($\phi'$ direction) and hard ($\beta_0'$ direction) polarization, which is along or orthogonal to the incident surface including the edge vector $\vec{\mathbf{e}}$. The factors $D_{0,n}^{\text{ISB,RSB}}$ are the coefficient to four different shadow boundaries with the detailed calculation in \cite{geng2013planungsmethoden}. 
Finally, the E-field strength after diffraction point $\boldsymbol{p}_{d}$ with distance $s$ is represented as
\begin{equation}
\begin{pmatrix}
\vec{E}_{\phi}^d(s) \\
\vec{E}_{\beta_0}^d(s)
\end{pmatrix}
=
\sqrt{\frac{s'}{s(s' + s)}} e^{-jk_0 s}
\begin{bmatrix}
-D_h & 0 \\
0 & -D_s
\end{bmatrix}
\begin{pmatrix}
\vec{E}_{\phi'}^i(\boldsymbol{p}_{d}) \\
\vec{E}_{\beta_0'}^i(\boldsymbol{p}_{d})
\end{pmatrix},
\label{dif_E}
\end{equation}
where $s'$ is the distance from $\boldsymbol{p}_{T}$ to $\boldsymbol{p}_{d}$, $\vec{E}_{\phi'}^i(\boldsymbol{p}_{d})$ and $\vec{E}_{\beta_0'}^i(\boldsymbol{p}_{d})$ are the incident E-field strength before $\boldsymbol{p}_{d}$. After diffraction, the soft ($\phi$ direction) and hard ($\beta_0$ direction) component is redefined according to the diffracted surface in Fig.~\ref{prop_phe}. With $\vec{E}_{\phi}^d(s)$ and $\vec{E}_{\beta_0}^d(s)$, a transformation into Cartesian coordinates is performed, and the components are summed to obtain $\vec{E}_{R,m}(x,y,z)$ for this diffraction path.

\begin{figure*}[!t]
\centering
\includegraphics[width=0.95\textwidth, trim=1mm 1mm 1mm 1mm,clip]{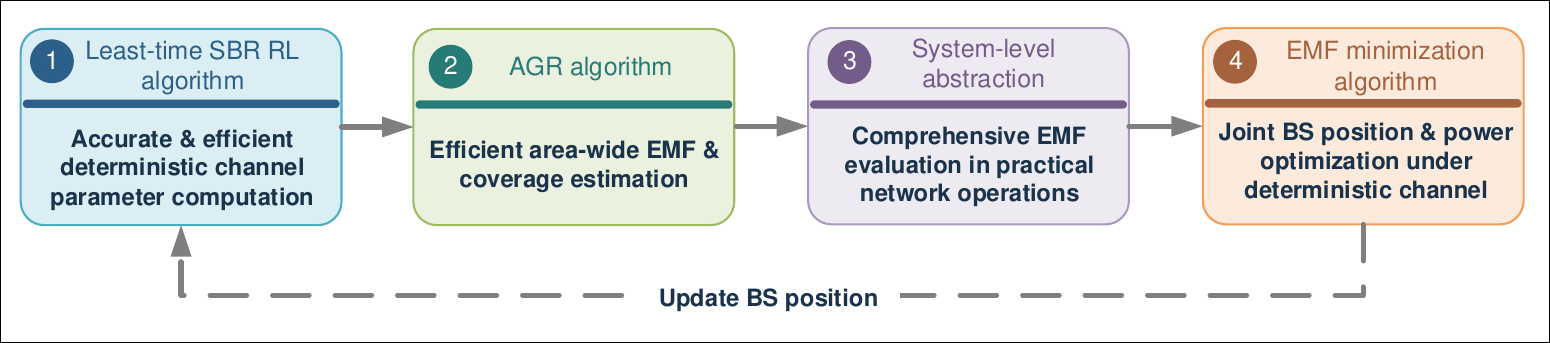}
\vspace{-1\baselineskip}
\caption{Overview of the proposed workflow, highlighting the purpose of each stage.}
\label{fig:workflow}
\end{figure*}

\section{Least-time \ac{SBR} \ac{RL} Algorithm and Real-world Measurement Validation}
\label{SBR_RL_algo}
Section~\ref{system_model} introduced the calculations for different propagation phenomena, with the determination of geometrical ray paths serving as the prerequisite. As the first stage of the proposed workflow in Fig.~\ref{fig:workflow}, a novel least-time \ac{SBR} algorithm is developed based on the NVIDIA OptiX framework \cite{parker2010optix} to identify the rays, enabling accurate and efficient deterministic channel modeling. In this section, the algorithm and its validation using real-world measurements are detailed.

\subsection{Least-time \ac{SBR} \ac{RL} Algorithm}
The least-time \ac{SBR} \ac{RL} algorithm aims to identify multipaths between \ac{Tx} and \ac{Rx} in a scenario similar to Fig.~\ref{prop_phe}. Before running into Algo.~\ref{alg:Simplified_Hybrid_SBR_RT}, the geometrical scenario must be imported, defined by each triangle mesh vertices $\mathcal{V}^* = \{\boldsymbol{v}_1,\boldsymbol{v}_2,\boldsymbol{v}_3\}$ and normals $\vec{\mathbf{e}}_n$ in 3D coordinate. The entire set of vertices and normals in the scenario can be grouped into $\mathcal{V}$ and $\mathcal{N}$, respectively. Based on $\mathcal{V}$, all the edges $\vec{\mathbf{e}}$ are obtained and saved in edge set $\mathcal{A}$. The inputs to Algo.~\ref{alg:Simplified_Hybrid_SBR_RT} consist of the geometry parameters, along with $\boldsymbol{p}_T$, $\boldsymbol{p}_R$ and other antenna configuration parameters mentioned in Section \ref{system_model}.

\begin{algorithm}
 \caption{Least-Time \ac{SBR} \ac{RL} Algorithm}
 \label{alg:Simplified_Hybrid_SBR_RT}
 \begin{algorithmic}[1]
 \renewcommand{\algorithmicrequire}{\textbf{Input:}}
 \renewcommand{\algorithmicensure}{\textbf{Output:}}
 \REQUIRE $\mathcal{V}$, $\mathcal{N}$, $\mathcal{A}$, $\boldsymbol{p}_T$, $\boldsymbol{p}_R$, $f_0$, $G_T$, $G_R$, $P_T$, $\vec{C}_R(\vartheta_{R}, \varphi_{R})$, $\vec{C}_T(\vartheta_T, \varphi_T)$, $M_{\text{dim}}$, $d_{\text{tr,max}}$
 \\ \textit{\ac{LoS} and Reflection Computation:}
  \STATE Initialize $s_{t,m} = 0$, $d_{\text{tr},m} = 0$, $\mathcal{V}_{t,m} = \emptyset$, for $m \in \{1,2, \dots, M_{\text{dim}}\}$
  \FOR{each $\vec{\mathbf{r}}_m \in \mathcal{R}$}
   \STATE Set $\boldsymbol{p}_{\text{O}} =  \boldsymbol{p}_T, \vec{\mathbf{e}}_d = \vec{\mathbf{r}}_m$
   \STATE $(\boldsymbol{p}_{h},\mathcal{V}^*,\vec{\mathbf{e}}_n) = \text{Tr}(\boldsymbol{p}_{\text{O}}, \vec{\mathbf{e}}_d)$
   \STATE $s_{t,m} = s_{t,m} + ||\boldsymbol{p}_{h}-\boldsymbol{p}_{\text{O}}||$
   \IF {$\boldsymbol{p}_{h} = \boldsymbol{p}_{R}$}
        \IF {$d_{\text{tr},m}=0$}
            \STATE Obtain $\vec{E}_{R,m}(x,y,z)$ with (\ref{E_field_around_T}) and $V_{R,m}$ with (\ref{U_R})
        \ELSIF{ $d_{\text{tr},m} \leq d_{\text{tr,max}}$} 
            \STATE Obtain $\vec{E}_{R,m}(x,y,z)$ with (\ref{modified_ref_E}) and $V_{R,m}$ with (\ref{U_R})
        \ENDIF
    \ELSE
        \STATE $d_{\text{tr},m} = d_{\text{tr},m}+1$, $\mathcal{V}_{t,m}  = \mathcal{V}_{t,m} \cup \mathcal{V}^*$
        \STATE Obtain $\vec{E}_{s}^i(\boldsymbol{p}_{r})$ and $\vec{E}_{p}^i(\boldsymbol{p}_{r})$ with (\ref{ref_E_1})
        \STATE Set $\vec{\mathbf{e}}_r = \vec{\mathbf{r}}_m - 2(\vec{\mathbf{r}}_m\cdot\vec{\mathbf{e}}_n)\vec{\mathbf{e}}_n$, $\boldsymbol{p}_{\text{O}} 
        = \boldsymbol{p}_{h}$, $\vec{\mathbf{e}}_d = \vec{\mathbf{e}}_r$ 
        \STATE go to line 4
    \ENDIF
  \ENDFOR
\FOR{each $\mathcal{V}_{t,m}$}
   \STATE $s_m = \min \{ s_{t,m'} \mid m' \in \{ m' \mid \mathcal{V}_{t,m'} = \mathcal{V}_{t,m} \} \}$
\ENDFOR
 \\ \textit{Diffraction Computation:}
 \FOR{each $\vec{\mathbf{e}}_{m} \in \mathcal{A}$}
    \STATE Calculate $\boldsymbol{p}_{d}$ 
        \STATE $\delta_{\text{O},1} = \text{Tr}_{\text{O}}(\boldsymbol{p}_T, \boldsymbol{p}_{d} - \boldsymbol{p}_{T})$, $\delta_{\text{O},2} = \text{Tr}_{\text{O}}(\boldsymbol{p}_{d}, \boldsymbol{p}_{R} - \boldsymbol{p}_{d})$
        \IF{$\delta_{\text{O},1} = \delta_{\text{O},2} = 0$}
            \STATE $s_{m} = ||\boldsymbol{p}_{d}- \boldsymbol{p}_{T}|| + ||\boldsymbol{p}_{d}- \boldsymbol{p}_{R}||$
            \STATE Obtain $\vec{E}_{R,m}(x,y,z)$ with (\ref{dif_E}) and $V_{R,m}$ with (\ref{U_R})
        \ENDIF
\ENDFOR
\ENSURE $\vec{E}_{R,m}(x,y,z)$, $V_{R,m}$, $\tau_m$, $(\varphi_{T,m},\vartheta_{T,m})$, and $(\varphi_{R,m},\vartheta_{R,m})$
 \end{algorithmic}
\end{algorithm}
The \ac{LoS} and reflection path determination in Algo.~\ref{alg:Simplified_Hybrid_SBR_RT} is based on the \ac{SBR} method \cite{didascalou2000ray}, where numerous rays are launched from the \ac{Tx} into the scenario and propagate through free space or undergo reflections, as described in lines 1-18. During the ray generation phase, sphere sampling ensures angularly uniform ray distribution across space. In total, \( M_{\text{dim}} \) rays, typically on the order of billions, are launched in line 2, with each unit ray direction labeled as \( \mathcal{R} = \{ \vec{\mathbf{r}}_1, \dots, \vec{\mathbf{r}}_m,..., \vec{\mathbf{r}}_{M_{\text{dim}}} \} \) with $m\in\{1,2,...,M_{\text{dim}}\}$. This is a notable difference that only a few representative rays are depicted in Fig.~\ref{prop_phe} for illustration. Before tracing process in line 3, the ray origin $\boldsymbol{p}_{\text{O}}$ is set to $\boldsymbol{p}_T$, and the ray direction is initialized using $\vec{\mathbf{r}}_m$. The tracing process is denoted as $\text{Tr}(\boldsymbol{p}_{\text{O}}, \vec{\mathbf{e}}_d)$ and the output is the hit point $\boldsymbol{p}_{h}$, $\mathcal{V}^*$ and $\vec{\mathbf{e}}_n$ of the intersected mesh in line 4. During this process, the total ray length $s_{t,m}$ is accumulated with $||\boldsymbol{p}_{h} - \boldsymbol{p}_{\text{O}}||$, which denotes their 3D distance in line 5. If a ray reaches $\boldsymbol{p}_R$ without obstruction following line 6–8 with the counter for reflections $d_{\text{tr},m}=0$ in line 7, it is a \ac{LoS} transmission path as in Fig.~\ref{prop_phe}. Subsequently, $\vec{E}_{R,m}(x,y,z)$ is calculated using (\ref{E_field_around_T}) in line 8. If a hit occurs other than $\boldsymbol{p}_R$ in line 12, $d_{\text{tr},m}$ accumulates in line 13 and ray reflects specularly in direction $\vec{\mathbf{e}}_r$ in line 15, calculated by $\vec{\mathbf{e}}_n$ and $\vec{\mathbf{e}}_m$. For multiple reflections, reflected ray is traced with $\boldsymbol{p}_{h}$ and $\vec{\mathbf{e}}_r$ in line 16 iteratively if $d_{\text{tr},m}$ is below the maximum $d_{\text{tr,max}}$ in line 9. If this ray ultimately reaches $\boldsymbol{p}_R$, the output parameters are calculated in line 10. In \ac{SBR}, \ac{Rx} is typically represented by a sphere whose radius significantly affects the number of collected rays. An excessively small sphere radius may cause certain rays to miss the sphere, while an overly large radius can capture an excessive number of rays, including redundant reflections from the same surface. 
To identify the unique path, \cite{didascalou2000ray} proposed procedure that consecutively test the number of reflections, time delay, and angles. The Algo.~\ref{alg:Simplified_Hybrid_SBR_RT} in this paper simplifies this by directly extracting the vertex $\mathcal{V}^*$ of intersected meshes in line 13, storing in $\mathcal{V}_{t,m}$ along the $m$-th ray whenever reflections occur. If the ray undergoes reflection(s) on the same mesh(es) in line 20, the shortest ray $s_m$ among them is selected, representing the unique reflection path. 
\begin{figure*}[!t]
\centering
\subfigure[]{\includegraphics[width=2.0in, trim=0mm 0mm 0mm 0mm,clip]
{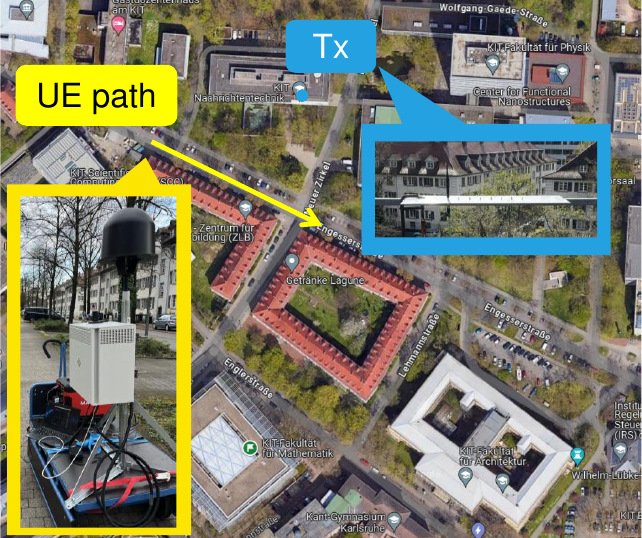}%
\label{measure_scene}
}
\subfigure[]{\includegraphics[width=2.0in, trim=0mm 0mm 0mm 0mm,clip]
{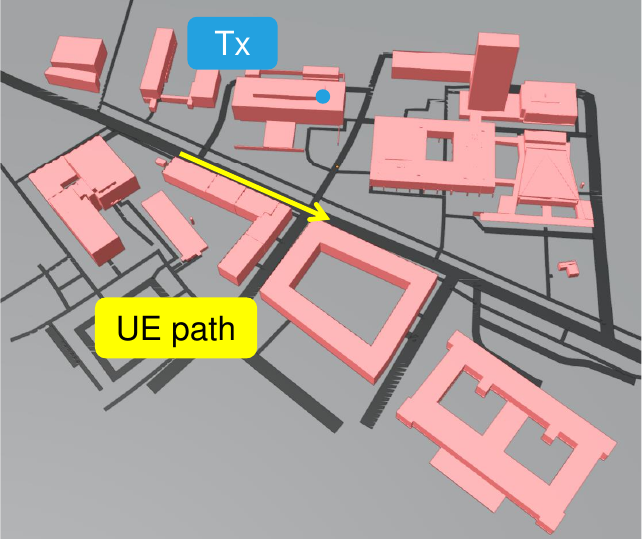}%
\label{sim_scene_1}
}
\subfigure[]{\includegraphics[width=2.5in, trim=0mm 0mm 0mm 0mm,clip]
{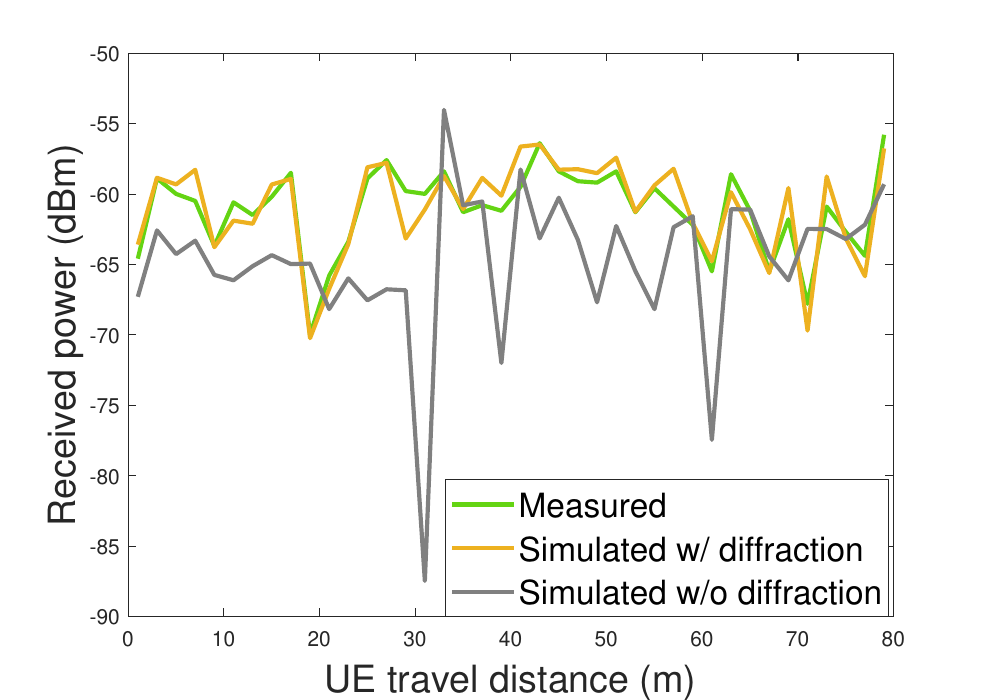}%
\label{measure_simu_results}
}
\caption{(a) Measurement scenario in Karlsruhe with an omni-directional \ac{Tx} antenna \raisebox{-0.5ex}{\color[rgb]{0.45,0.75,1}\Large \textbullet} placed on the rooftop at a height of \SI{19.18}{m}, transmitting a continuous signal with $P_T = \SI{10}{dBm}$, $G_T = \SI{5}{dBi}$, at $f_0 = \SI{2.5}{GHz}$. The \ac{Rx} is UMS400 universal monitoring system from Rohde \& Schwarz \cite{rohde_schwarz} with $G_R = \SI{10}{dBi}$ at height \SI{2}{m} and is located at different positions along {\color{yellow!75!orange}\Large$\rightarrow$} with \SI{2}{m} intervals. (b) Simulation scenario with $M_{\text{dim}} = 2.5$ billion and \ac{Rx} sphere radius of \SI{0.24}{m}. The $d_{\text{tr,max}}$ is set to 5, and the maximum number of diffraction is 2. (c) Comparison of received power between measurement and simulation. }
\vspace{-1\baselineskip}
\label{fig_2}
\end{figure*} 
Continuing to use the \ac{SBR} method to identify diffraction points, as in \cite{9753133}, is an intuitive approach. However, it introduces two main issues. First, determining whether a ray intersects with an edge necessitates an approximate range due to the limited number of rays, which in turn reduces accuracy. Second, when considering the wedge diffraction based on Keller’s cone \cite{keller1962geometrical}, diffracted rays form a cone, adhering to the law that the incident angle equals the diffraction angle. This requires treating the diffraction point as a secondary source, emitting a bunch of rays that fulfill the law. The implementation is computationally complex, especially when higher-order diffraction is included, significantly increasing the number of required rays.
Therefore, Fermat's principle is applied to identifying the diffraction points, which states that the ray from $\boldsymbol{p}_{T}$ to $\boldsymbol{p}_{R}$, passing through one or more diffraction points, must be the shortest \cite{geng2013planungsmethoden}. Taking first-order diffraction as an example, the $\boldsymbol{p}_{d}$ for each edge $\vec{\mathbf{e}}_m\in \mathcal{A}$ is calculated in line 23, by considering the \ac{Tx}, the edge surfaces, and the \ac{Rx} point neglecting other objects, as illustrated in Fig.~\ref{prop_phe}. Since calculating $\boldsymbol{p}_{d}$ analytically is impractical in C++ environment, it is determined numerically using a combination of golden search and Newton's method \cite{fugen2006capability}. Subsequently, all objects in the scenario are considered, and it is verified that $\boldsymbol{p}_{d}$ is accessible from both $\boldsymbol{p}_{T}$ and $\boldsymbol{p}_{R}$ without obstruction. This is achieved by tracing the occlusion ray in line 24, represented by $\text{Tr}_{\text{O}}(\boldsymbol{p}_{\text{O}}, \vec{\mathbf{e}}_d)$ with the output $\delta_{\text{O}} = 0$ to indicate the absence of occlusion in line 25. Afterwards, output parameters are calculated for this diffraction path in line 26 and 27. In addition to $\vec{E}_{R,m}(x, y, z)$ and $V_{R,m}$, time delay is derived from the total ray length as $\tau_m = s_m / c_0$, and the angles are determined using the first and last hit point. 

The calculation efficiency of the least-time \ac{SBR} \ac{RL} algorithm can be attributed to several key factors. First, the geometrical scenario is optimized using the \ac{GAS} in the OptiX framework, which accelerates \ac{RL} by efficiently organizing geometry to reduce intersection tests. Second, the parallel processing capabilities of the GPU are leveraged. Third, Algo.~\ref{alg:Simplified_Hybrid_SBR_RT} applies Fermat's principle to identify unique reflection and diffraction paths, significantly lowering computational effort and memory usage. Lastly, the \ac{Rx} sphere is hard-coded and separate from geometry, enabling simulations with multiple \ac{Rx} positions by simply updating its position without reimporting geometry scenario, thereby reducing computation time.

\subsection{\ac{RL} Validation through Real-world Measurement}
\label{validation}
Given the critical need for accuracy in assessing \ac{EMF} exposure, the \ac{RL} tool must be validated through real-world measurements to ensure its precision. An aerial view of the measurement scenario is provided in Fig.~\ref{measure_scene} with all the measurement settings in the caption. The 3D simulation scenario imported from \ac{OSM}, as shown in Fig.~\ref{sim_scene_1}, which includes all the building structures in the area. The parameters, such as $f_0$, $P_T$, $G_T$, $G_R$, $\boldsymbol{p}_T$ and $\boldsymbol{p}_R$ are all consistent with the measurement settings. Previous simulators at millimeter wave frequency omitted diffraction \cite{felbecker2012electromagnetic}, given its reduced significance at those frequencies. In our simulation, when diffraction is not accounted for, the \ac{RMSE} between the simulated and measured results is \SI{6.63}{dB}, with their received power depicted by the gray and green curves in Fig.~\ref{measure_simu_results}. By incorporating diffraction, as represented by the yellow curve in Fig.~\ref{measure_simu_results}, the accuracy significantly improves with reduced \ac{RMSE} to \SI{1.29}{dB}, highlighting the importance of considering diffraction. 
Normally, the received signal power is measured for each received packet in smartphones quantified by \ac{RSSI} and a typical commercially available transceiver is expected to provide \SI{4}{dB} \ac{RSSI} accuracy \cite{asaad2022comprehensive}. Therefore, this minor \ac{RMSE} compared to \ac{RSSI} accuracy demonstrates that Algo.~\ref{alg:Simplified_Hybrid_SBR_RT} is sufficiently accurate for estimating the received power. Another advantage of the proposed Algo.~\ref{alg:Simplified_Hybrid_SBR_RT} is the calculation efficiency. With one pair of \ac{Tx} and \ac{Rx} in the simulation scenario with 13,963 polygons, the \ac{RL} algorithm requires \SI{7}{s}, compared to 45 minutes using \ac{RT} in \cite{fugen2006capability}. 
The accuracy and efficiency of Algo.~\ref{alg:Simplified_Hybrid_SBR_RT} support repeated deterministic channel simulations for different \ac{Tx} and \ac{Rx} position pairs in urban areas, providing a basis for evaluating EMF exposure and coverage in subsequent BS deployment optimization.

\section{\ac{EMF} Exposure and Signal Coverage Estimation over Target Area}
\label{agr_algo}
Up to this point, \ac{EMF} exposure and received power can be calculated at an arbitrary position in the 3D urban scenario using Algo.~\ref{alg:Simplified_Hybrid_SBR_RT}. As the second stage of the proposed workflow in Fig.~\ref{fig:workflow}, to extend the calculation efficiently over the entire target area, the \ac{AGR} algorithm is introduced in this section, enabling global estimation of \ac{EMF} exposure and coverage. It exploits the spatial stability of the effective channel while accounting for BS beamforming, reducing the computational effort of exhaustive evaluation on uniformly fine grids.

\subsection{Channel Model with Beamforming}
\label{sec_beamforming}
Beamforming at the BS should be incorporated since the concentrated power influences the EMF exposure for UE. Assuming power matching at the input of the \ac{Tx} antenna, the transmitted amplitude $V_T$ is represented as $V_T = \sqrt{8 \Re(Z^*_{AT}) P_T}$ with $Z^*_{AT}$ the complex conjugate impedance of the \ac{Tx} antenna. Then, the channel frequency response $h(f)$ is calculated from the ratio of $V_R$ in (\ref{P_R}) and $V_T$
\begin{equation}
\label{channel_response}
\begin{aligned}
h(f) = &\frac{V_R}{V_T} = \frac{V_R}{\sqrt{8 \Re (Z_{AT}^*) P_T}}\\
=&\sqrt{{(\frac{\lambda_{0}}{4\pi})}^2 G_T G_R}\cdot \sum_{m=1}^{M_p} \vec{C}_{R}(\vartheta_{R,m},\varphi_{R,m}) \\
 & \cdot T_{m} \cdot \vec{C}_{T}(\vartheta_{T,m},\varphi_{T,m}) e^{-j k_0 s_{m}} 
\end{aligned}
\end{equation}
where $T_{m}$ is the full polarimetric transmission matrix along $m$-th path. Now considering \ac{Tx} as BS equipped with $M_{\text{ant}}$ antennas and the \ac{Rx} as \ac{UE} with a single omni-directional antenna, the channel matrix is $\mathbf{H} \in \mathbb{C}^{1 \times M_{\text{ant}}}$, with each element computed using (\ref{channel_response}) between $M_{\text{ant}}$ BS antennas and the \ac{UE} antenna.
When digital beamforming is performed, the effective complex channel coefficient including digital beamforming at the BS is written as
\begin{equation}
h_{\text{bf}} = \mathbf{H} \boldsymbol{w}_{b,u},
\label{F_BF}
\end{equation}
where $\boldsymbol{w}_{b,u} \in \mathbb{C}^{M_{\text{ant}}\times 1}$ is the beamforming vector at the BS to the \ac{UE}. Assuming that \ac{CSI} is available at the BS through a backhaul link, \ac{MRT} is implemented as a typical beamforming technique since it maximizes the \ac{SNR}. Please notice that other beamforming schemes can be also applied, which shares the same optimization procedure therefore not discussed for simplicity.

\subsection{Estimation over Target Area using \ac{AGR} Algorithm}
\label{ARG}
A common approach is to discretize the target area into uniform fixed-size grids and obtain the \ac{EMF} exposure at each grid center position with virtual \ac{UE}. This procedure normally assumes that the large-scale fading of the channel remains similar inside the grids, as done in many studies \cite{amaldi2003planning}, \cite{khalek2011optimization}, \cite{siomina2006automated}, \cite{sun2018jointly} and \cite{11737699}, etc. However, the reason for the chosen grid size or the exact value of it is usually not specified, as many studies typically consider scenarios without any geometrical building and adopt empirical channel models. In fact, an excessively large grid size may lead to diminished accuracy, while overly refining the grid size, though enhancing precision, can substantially increase computation complexity. In a practical urban scenario, the selection of grid size is crucial because the channel conditions at a road intersection, where \ac{LoS} paths are likely to dominate, must differ significantly from the case on a street flanked by buildings, where numerous \ac{NLoS} paths prevail. Therefore, the goal is to reduce computational complexity while maintaining high accuracy by not only determining an appropriate grid size but also adapting it flexibly to varying channel conditions in an urban scenario. Considering the factors, \ac{AGR} is developed as a potential solution, which is realized by reducing the grid size when significant changes in the channel are detected.

\begin{figure*}[!t]
\centering
\includegraphics[width=5.5in, trim= 3mm 1mm 1mm 3mm,clip]{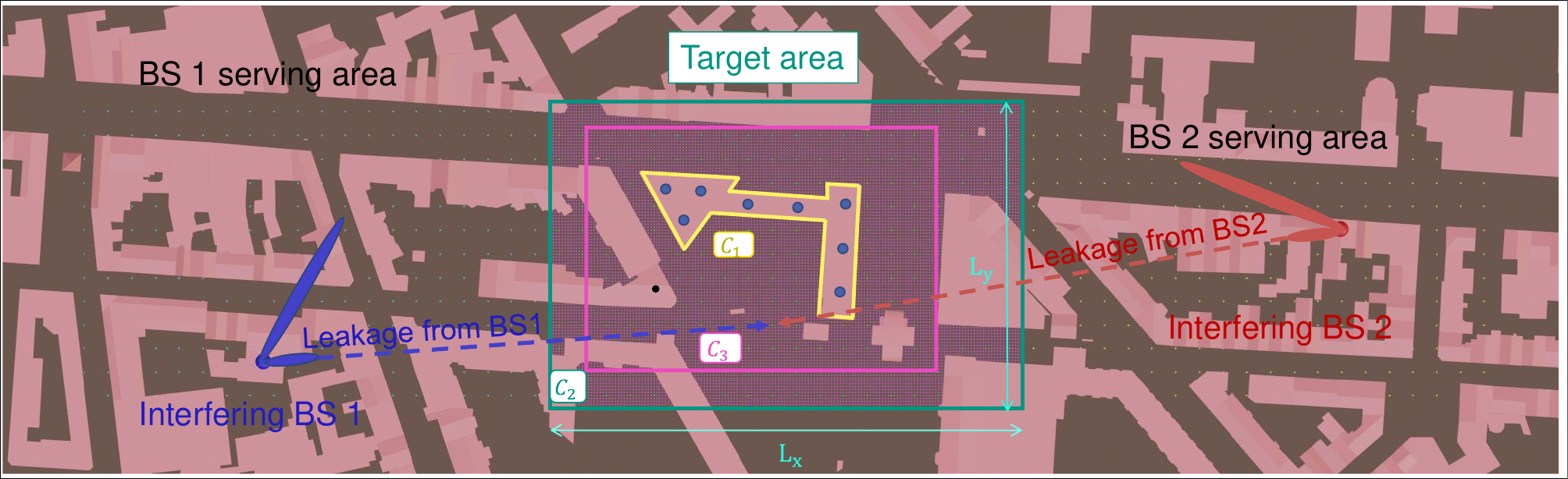}
\caption{Aerial view of 3D simulation scenario. Two interfering BSs are marked by spheres \raisebox{-0.5ex}{\textcolor[RGB]{0,0,255}{\Large\textbullet}} and \raisebox{-0.5ex}{\textcolor[RGB]{255,0,0}{\Large\textbullet}}, serving their own UEs represented as \textcolor[RGB]{74,190,202}{\Large\textbullet} and \textcolor[RGB]{255,255,0}{\Large\textbullet}, respectively. In the middle is the target area with different boundary definition. The grid center positions \raisebox{-0.5ex}{\color[rgb]{0,0.8,0}\Large \textbullet} form $\mathcal{U}_1$ with initial grid length $g_1$, \raisebox{-0.5ex}{\color[rgb]{0.55,0,0.8}\Large \textbullet} form the final $\mathcal{U}_0$ with minimum grid length $g_0$. The initial $\mathcal{B}^0$ consists of \raisebox{-0.5ex}{\textcolor[RGB]{66,92,155}{\Large \textbullet}} and position origin is \raisebox{-0.5ex}{\color[rgb]{0,0,0}\Large \textbullet}.}
\label{bound_def}
\end{figure*}

In practical cases, the target area is an urban scenario similar to Fig.~\ref{bound_def}, with all the grid center positions are set at the same height, which reduces the problem to 2D estimation. Assuming the square-shaped grid, the grid size is thus simplified to the grid length. The inputs to Algo.~\ref{algo_2} are minimum grid length $g_{0}$, an initial grid length $g_1$, $\boldsymbol{p}_T$, and $\mathcal{U}_1$ set. A unit \ac{EIRP} is assumed and thus not specified at input of Algo.~\ref{algo_2}, since it does not affect grid refinement process and allows directly scaling to the applied \ac{EIRP}. The set $\mathcal{U}_1$ consists of grid center positions $\boldsymbol{u}_{(i,j)}^1$ distributed uniformly with grid length $g_1$ inside the target area boundary $C_2$ with \(i \in \{1, 2, \ldots, \lceil \frac{L_x}{g_1} \rceil\}\) and \(j \in \{1, 2, \ldots, \lceil \frac{L_y}{g_1} \rceil\}\), where length $L_x$ and width $L_y$ are the dimension of $C_2$ along the $x$ and $y$ axes, respectively. This initial $g_1$ is chosen as the smallest street width in the target area, ensuring at least one grid center for any street. Additionally, $\mathcal{U}_{\text{tot}}$ represents the total set of grid center positions, which needs to be simulated using Algo.~\ref{alg:Simplified_Hybrid_SBR_RT}, starting with $\mathcal{U}_{\text{tot}} = \mathcal{U}_1$ in line 1. Then, considering beamforming in (\ref{F_BF}), the \ac{EMF} exposure and received power at grid center position $\boldsymbol{u}_{(i,j)}^k$ are calculated in $k$-th iteration, denoting as $E_{\text{bf}}(\boldsymbol{u}_{(i,j)}^k)$ and $P_{R,\text{bf}}(\boldsymbol{u}_{(i,j)}^k)$ in line 3, respectively. 
If a substantial received power difference between $\boldsymbol{u}_{(i,j)}^k$ and any of its neighboring positions in $\mathcal{V}_{\text{nb}}$, defined in line 6 and 8, exceeds the threshold $\Delta P_{R,\text{max}}$ in line 11, additional simulations are required. Specifically, intermediate positions are added to $\mathcal{U}_{k}$ for further simulations in line 13, yielding a refined grid length $g_{k}$ in line 17. The $\mathcal{U}_{\text{tot}}$ is updated by incorporating $\mathcal{U}_k$ in line 18, and the refinement proceeds until $g_{0}$ is reached or no additional positions are required in line 2. After obtaining the complete $\mathcal{U}_{\text{tot}}$, $E_\text{bf}(\boldsymbol{u}_{\text{tot}})$ and $P_{R,\text{bf}}(\boldsymbol{u}_{\text{tot}})$ are derived, where $\boldsymbol{u}_{\text{tot}}$ represents each element in $\mathcal{U}_{\text{tot}}$. Finally, 2D linear interpolation is performed in line 20 to complete the received power and \ac{EMF} exposure estimation for minimum grid length $g_{0}$, yielding $E_\text{bf}^{\text{unit}}(\boldsymbol{u})$ and $P^{\text{unit}}_{R,\text{bf}}(\boldsymbol{u})$, where $\boldsymbol{u}\in \mathcal{U}_0$ and $\mathcal{U}_0$ is the set of grid center positions with minimum grid length $g_0$. The superscript emphasizes that the estimations are derived under unit \ac{EIRP} in Algo.~\ref{algo_2}.

In summary, by exploiting the spatial stability of the effective channel while accounting for BS beamforming, the AGR algorithm efficiently discretizes the target area while balancing computational effort and estimation accuracy. The resulting EMF exposure and received power maps provide a basis for evaluating different BS deployments.

\begin{algorithm}
 \caption{\ac{EMF} exposure estimation using \ac{AGR} algorithm}
 \begin{algorithmic}[1]
 \renewcommand{\algorithmicrequire}{\textbf{Input:}}
 \renewcommand{\algorithmicensure}{\textbf{Output:}}
 \REQUIRE $g_0$, $g_1$, $\boldsymbol{p}_T$, $\mathcal{U}_1$
  \\ \textit{Initialization}:
  \STATE $k=1$, $\mathcal{U}_{\text{tot}} = \mathcal{U}_1$
  \WHILE{ $\frac{g_1}{2^k} > g_{0}$ or $\mathcal{U}_{k} \neq \emptyset$} 
  \STATE Obtain $E_\text{bf}(\boldsymbol{u}_{(i,j)}^k)$ and $P_{R,\text{bf}}(\boldsymbol{u}_{(i,j)}^k)$ with $\boldsymbol{u}_{(i,j)}^k \in \mathcal{U}_k$
 \FOR {each \(\boldsymbol{u}_{(i,j)}^k \in \mathcal{U}_k\)}
    \IF{$k > 1$}
        \STATE $\mathcal{V}_{\text{nb}} = \{\boldsymbol{u}_{(i+\frac{1}{2^k},j)}^{k-1}, \boldsymbol{u}_{(i-\frac{1}{2^k},j)}^{k-1}, \boldsymbol{u}_{(i,j+\frac{1}{2^k})}^{k-1}, \boldsymbol{u}_{(i,j-\frac{1}{2^k})}^{k-1}\}$
    \ELSE
        \STATE $\mathcal{V}_{\text{nb}} = \{\boldsymbol{u}_{(i+1,j)}^k, \boldsymbol{u}_{(i-1,j)}^k, \boldsymbol{u}_{(i,j+1)}^k, \boldsymbol{u}_{(i,j-1)}^k\}$
    \ENDIF
    \FOR {each point \(\boldsymbol{u}_{(i',j')}^k \in \mathcal{V}_{\text{nb}}\)}
        \IF{$|P_{R,\text{bf}}(\boldsymbol{u}_{(i,j)}^k) - P_{R,\text{bf}}(\boldsymbol{u}_{(i',j')}^k)| > \Delta P_{R,\text{max}}$}
            \STATE $k=k+1$
            \STATE Add $\boldsymbol{u}_{(\frac{i+i'}{2}, \frac{j+j'}{2})}^k$ into $\mathcal{U}_k$
        \ENDIF
    \ENDFOR
\ENDFOR
    \STATE Refine grid by: $g_k = g_{k-1}/2$
    \STATE Update $\mathcal{U}_{\text{tot}} = \mathcal{U}_{\text{tot}} \cup \mathcal{U}_k$
  \ENDWHILE
   \STATE Perform 2D linear interpolation for $P_{R,\text{bf}}(\boldsymbol{u}_{\text{tot}})$, $E_{\text{bf}}(\boldsymbol{u}_{\text{tot}})$ for each $\boldsymbol{u}_{\text{tot}} \in \mathcal{U}_{\text{tot}}$
\ENSURE $E^{\text{unit}}_\text{bf}(\boldsymbol{u})$ and $P^{\text{unit}}_{R,\text{bf}}(\boldsymbol{u})$ for each $\boldsymbol{u} \in \mathcal{U}_0$
 \end{algorithmic} 
 \label{algo_2}
\end{algorithm}

\section{Problem Formulation and Proposed Solution}
\label{BS_place_problem_and_NM}
After obtaining the global \ac{EMF} exposure and received power, the workflow incorporates system-level abstraction as its third stage in Fig.~\ref{fig:workflow}, enabling a more comprehensive assessment of \ac{EMF} exposure under practical network operation. In this section, in addition to the beamforming implemented in (\ref{F_BF}), the BS side incorporates the actual transmit power, intercell interference and \ac{CSI} imperfection, while on the \ac{UE} side, mobility is modeled over the \ac{EMF} exposure averaging interval. Building on the preceding procedures, the fourth stage in Fig.~\ref{fig:workflow} formulates the coverage-guaranteed EMF exposure minimization problem and solves it using a geometry-aware algorithm that enables joint optimization of BS deployment and power based on deterministic channel simulations.

\subsection{Practical Network Behaviors Consideration}
To represent practical network behaviors, the following aspects are considered.
\subsubsection{Actual Transmit Power}\label{actual maximum}
Accounting for the fact that a BS does not transmit at maximum power continuously, incorporating actual transmit power is necessary to avoid overestimation of time-averaged \ac{EMF} exposure. Given that instantaneous transmit power in practical BSs varies with traffic load, scheduling, and power control, the actual maximum approach is developed to obtain a statistical estimate of time-averaged BS transmit power \cite{xu2021actual}, following the International Electrotechnical Commission (IEC) 62232:2017 method \cite{IEC62232_2017}. 
Based on the deterministic factors such as duty cycle and stochastic factors derived from \ac{CDF} of the time-averaged transmit power, the \ac{PRF} describes the ratio of the actual transmit power $P_{T,\text{act}}$ to the theoretical maximum $P_{T,\text{max}}$ \cite{thors2017time}, \cite{chiaraviglio2021health}
\begin{equation}
   P_{\text{RF}} =  P_{T,\text{act}}/P_{T,\text{max}}.
\end{equation}
Due to the fact that $P_{\text{RF}}$ depends on the total number of connected \ac{UE}s $N$ \cite{10500988}, it is denoted as $P_{\text{RF}}(N)$, leading to $P_{T,\text{act}}(N)$.
Based on the output of Algo.~\ref{algo_2} by setting the input $\boldsymbol{p}_T$ to the BS position $\boldsymbol{p}_b=[x_b,y_b,z_b]$, the EMF exposure and received power using the actual \ac{EIRP} with $P_{T,\text{act}}(N)G_T$ are obtained as $E_{\text{bf}}(\boldsymbol{\zeta},\boldsymbol{u},N)$ and $P_{R,\text{bf}}(\boldsymbol{\zeta},\boldsymbol{u},N)$, where $\boldsymbol{\zeta} = (\boldsymbol{p}_b,EIRP)$ collects the optimization variables with $EIRP = P_{T,\text{max}}G_T$, representing theoretical maximum \ac{EIRP}. 

\subsubsection{Intercell Interference}
Frequency reuse across neighboring cells inevitably introduces intercell interference. The consideration of intercell interference yields an accurate EMF exposure evaluation, as it degrades \ac{SINR} while increasing total field strength at the \ac{UE}. In practical networks, interfering BSs dynamically steer their beams towards their own UEs as in Fig.~\ref{bound_def}, and the resulting temporal and spatial variations in leakage power are modeled using a procedure based on the 3GPP specifications \cite{atis38828_2019}, which is also adopted in our previous study \cite{11368706}.
Consider the $l$-th interfering BS at $\boldsymbol{p}_l=[x_l,y_l,z_l]$, serving a UE at $\boldsymbol{u}_l\in\mathcal{U}_l$, where $\mathcal{U}_l$ denotes the set of UE positions in its serving area in Fig.~\ref{bound_def}. The corresponding channel vector $\mathbf{H}_l(\boldsymbol{u}_l)$ is obtained using \eqref{channel_response}, and the effective complex channel coefficient including beamforming is determined based on (\ref{F_BF}) as
\begin{equation}
    h_{\text{bf},l}(\boldsymbol{u}_l) = \mathbf{H}_{l}(\boldsymbol{u}_l)\boldsymbol{w}_{l}(\boldsymbol{u}_l),
\end{equation}
where $\boldsymbol{w}_l(\boldsymbol{u}_l)$ is the BS beamforming vector. When an interfering BS transmits to its own UE, the resulting leakage power to the UE position $\boldsymbol{u}$ in the target area can be expressed as
\begin{equation}
\label{P_RL}
P_{R,l}(\boldsymbol{u},\boldsymbol{u}_l)
=
P_{\mathrm{T},l}
\left|
\mathbf{H}_l(\boldsymbol{u})
\boldsymbol{w}_l(\boldsymbol{u}_l)
\right|^2,
\end{equation}
where $P_{\mathrm{T},l}$ denotes the transmit power of the $l$-th interfering BS. Considering that the served UE is equally likely to occupy any position in $\mathcal{U}_l$ and each UE is allocated the same service duration under round-robin scheduling as adopted in \cite{atis38828_2019}, the average interference power from the $l$-th interfering BS is
\begin{equation}
\label{take_avg}
P_{R,l}^{\mathrm{avg}}(\boldsymbol{u})
=
\frac{1}{|\mathcal{U}_l|}
\sum_{\boldsymbol{u}_l\in\mathcal{U}_l}
P_{R,l}(\boldsymbol{u},\boldsymbol{u}_l), 
\end{equation}
where $|\cdot|$ represents the number of elements in the set. Then, the average interference power in \eqref{take_avg} is evaluated across the entire target area at the finest grid resolution using the AGR algorithm in Algo.~\ref{algo_2} with $\boldsymbol{p}_T=\boldsymbol{p}_l$, yielding $P_{R,l}^{\mathrm{avg}}(\boldsymbol{u})$ for each $\boldsymbol{u}\in\mathcal{U}_0$.
Finally, the \ac{SINR} for the $n$-th UE is calculated as
\begin{equation}
\label{sinr_ideal}
 \gamma^{\text{ideal}}(\boldsymbol{\zeta}, \boldsymbol{u}, N) = \frac{P_{R,\text{bf}}(\boldsymbol{\zeta}, \boldsymbol{u}, N)}{P_{R,L}(\boldsymbol{u}) + \sigma_n^2} = \frac{P_{R,\text{bf}}(\boldsymbol{\zeta}, \boldsymbol{u}, N)}{\sum_{l=1}^{L} P_{R,l}^{\text{avg}}(\boldsymbol{u}) + \sigma_n^2},
\end{equation}
where $P_{R,L}(\boldsymbol{u})$ is the total interference power and $\sigma_n^2$ is the noise power for the $n$-th UE. 

\vspace{2pt}
\subsubsection{\ac{CSI} Imperfection}
Imperfect \ac{CSI} degrades received signal quality and to compensate for this, the transmit power of the desired signal should be increased, thereby elevating \ac{EMF} exposure. In practice, \ac{CSI} is obtained by channel estimation and is commonly modeled with an additive independent random error term. Under perfect channel reciprocity, the linear \ac{SINR} reduction factor between the imperfect CSI and the perfect CSI is expressed as \cite{mi2017massive}
\begin{equation} \label{gamma_rf}
    \delta_{\text{CSI}} = 1-\tau_e^2,
\end{equation}
where $\tau_e^2$ is estimation variance and $\tau_e = 0$ implies perfect estimation.
Therefore, the \ac{SINR} considering an imperfect CSI for each grid center is $\gamma(\boldsymbol{\zeta}, \boldsymbol{u}, N)=  \delta_{\text{CSI}}\gamma^{\text{ideal}}(\boldsymbol{\zeta}, \boldsymbol{u}, N)$ based on (\ref{sinr_ideal}).

\subsubsection{UE Mobility} \label{UE mobility}
To capture realistic \ac{UE} dynamics in urban scenarios and quantify the experienced EMF exposure, \ac{UE} mobility over the \ac{EMF} exposure averaging interval is incorporated.
In comparison to the agent-based model in \cite{leeman2025city}, the \ac{UE} mobility model proposed in this paper captures pedestrian dynamics in the urban environment more comprehensively. The model is based on correlated random walk (CRW) with von Mises distributed heading increments \cite{turchin1998quantitative}, that captures typical pedestrian behaviors, including wandering, stochastic pausing, crossing streets, and window shopping. The urban environment is separated into building, street and pedestrian area (also see Fig.~\ref{P2_opt}). 
At each time frame, \ac{UE} advances at a nominal walking speed $v$ with a heading update $\Delta\theta$ drawn from a von Mises distribution with concentration $\kappa$, which models randomized tendencies of different \ac{UE}s to walk in a straight line. Pausing behavior is included via a stop probability and a log-normally distributed pause duration, yielding realistic dwell times. The \ac{UE} is not allowed to enter building area, as indoor \ac{EMF} exposure is substantially lower. However, it may walk along the building facades, mimicking window shopping behavior.
With the entry probability $\rho_{\text{entry}}$, \ac{UE} is permitted to enter the street and retains its current heading, thereby traversing the street in a straight path instead of reorienting. Therefore, the \ac{UE} mobility model conforms the generated trajectories to the urban topology.

To evaluate the \ac{EMF} exposure for a dynamic \ac{UE} over the \ac{EMF} exposure averaging interval $T_{\text{avg}}$, precomputed global \ac{EMF} exposure and received power serve as lookup tables queried along its trajectory. At each time frame $t = \{1,2,...,T\}$ with $T = T_{\text{avg}}/T_{\text{step}}$ denoting the total number of time frames with time step $T_{\text{step}}$, the $n$-th \ac{UE} position is mapped to its nearest grid center position $\boldsymbol{u}_n(t)\in \mathcal{U}_0$, and the corresponding $E_{\text{bf}}(\boldsymbol{\zeta},\boldsymbol{u}_n(t),N)$ and $P_{R,\text{bf}}(\boldsymbol{\zeta},\boldsymbol{u}_n(t),N)$ are obtained.
Along the trajectory of $n$-th \ac{UE}, let $N_{\text{occ}}(\boldsymbol{u}_n(t)) \leq N$ denote the number of \ac{UE}s that are mapped to the same grid center position $\boldsymbol{u}_n(t)$ at $t$-th time frame. Co-location of multiple \ac{UE}s increases \ac{EMF} exposure, and under worst-case condition the total \ac{EMF} exposure for the $n$-th UE at $t$-th time frame is given by
\begin{align} \label{tot emf}
    &E_{\text{bf,tot}}(\boldsymbol{\zeta},\boldsymbol{u}_n(t),N) = \notag\\
    &\quad \sqrt{\frac{Z_{F0}}{A_e}[ P_{\text{R},L}(\boldsymbol{u}_n(t))\! +\!N_{\text{occ}}(\boldsymbol{u}_n(t))P_{R,\text{bf}}(\boldsymbol{\zeta},\boldsymbol{u}_{n}(t),N)]},
\end{align}
where $A_e= \lambda_0^2G_R/4\pi$ is the effective antenna aperture. This expression further includes the additional interference power under the assumption that the target and interfering BSs signals are statistically independent.
Therefore, the \ac{RMS} time-averaged \ac{EMF} exposure for the $n$-th \ac{UE} is
\begin{equation} \label{EMF avg for one UE}
    \bar{E}_n(\boldsymbol{\zeta},N) = 
    \sqrt{\frac{1}{T}\sum_{t=1}^{T}E_{\text{bf,tot}}^2(\boldsymbol{\zeta},\boldsymbol{u}_n(t),N)}.
\end{equation}

\subsection{Coverage-guaranteed EMF Exposure Minimization}
\label{BS_place_problem}
Upon the aforementioned aspects, the objective is to minimize \ac{EMF} exposure while maintaining coverage by jointly optimizing BS deployment and radiated power. Since the variability in number of \ac{UE}s and the randomness in \ac{UE} trajectories render this objective stochastic, \ac{MC} simulation is employed to provide a statistically consistent estimate of the time-averaged \ac{EMF} exposure in (\ref{EMF avg for one UE}), written as
\begin{equation} \label{MC avg}
    J(\boldsymbol{\zeta}) = \mathbb{E}\left[ \frac{1}{N(\omega)} \sum_{n=1}^{N(\omega)}\bar{E}_n(\boldsymbol{\zeta},N(\omega))\right],
\end{equation}
where $\mathbb{E}[\cdot]$ represents the expectation of the argument and $N(\omega)$ is the number of \ac{UE}s in $\omega$-th \ac{MC} simulation. To ensure a stable mean estimate in (\ref{MC avg}), the sample mean and variance after each \ac{MC} run are updated and a fixed-width 95\% confidence interval stopping rule was applied, terminating when the half-width falls below a small tolerance $\delta_{\text{tol}}$. 
Finally, the coverage-guaranteed EMF exposure minimization problem is formulated as
\begin{equation}
\tag{P0}
    \operatorname*{argmin}_{\boldsymbol{\zeta}} J(\boldsymbol{\zeta})
    \label{P0}
\end{equation}
subject to:
\begin{equation} \label{BS limits}
    \boldsymbol{p}_b \in C_1,
\end{equation}
\begin{equation}
   \frac{|\mathcal{U}_{\text{cov}}|}{|\mathcal{U}_0|} \geq \rho_{\text{cov}}, 
   \label{cov_c}
\end{equation}
\begin{equation}
   EIRP < EIRP_{\text{max}}.
   \label{EIRP}
\end{equation}
The feasible region of BS deployment is constrained as in (\ref{BS limits}) on the rooftop of the center building, which has a contour $C_1$ as shown in Fig.~\ref{bound_def} due to practical limitations. Typically, network providers face restrictions for BS site leasing or permission, which limits the available area for BS deployment. Coverage is defined as the percentage of grid center positions in $\mathcal{U}_0$ that satisfy the minimum SINR threshold $\gamma_{\text{min}}$ with \( \mathcal{U}_{\text{cov}}= \{\boldsymbol{u} \in \mathcal{U}_0|\gamma(\boldsymbol{\zeta}, \boldsymbol{u},N) \geq \gamma_{\text{min}}\} \), and $\rho_{\text{cov}}$ is the required coverage in (\ref{cov_c}). To guarantee proper coverage, the \ac{EIRP} of the BS is adjusted, while respecting a regulatory limit $EIRP_{\text{max}}$ as in (\ref{EIRP}).

\subsection{Proposed Solution based on \ac{NM} Method}
To identify a suitable solution for problem (\ref{P0}), its inherent complexities should be analyzed. Firstly, the constraint implied in (\ref{BS limits}) is typically non-convex since the contour of the building might not be convex. Secondly, wave propagation in urban environments produces sharp variations due to different channel conditions, therefore the mapping from optimization variables to the objective is non-smooth and non-differentiable. Even a small change to $\boldsymbol{p}_b$ can result in significant changes to the objective value, as this highly depends on the scenario geometry. Finally, stochastic \ac{UE} trajectories, grid discretization and discrete coverage definition all introduce discontinuities, making analytic gradients or even a finite-difference approximation unavailable. To address this problem, the non-convex constraint in problem (\ref{P0}) is first transferred into a convex form with a penalty factor, and the resulting problem is solved based on the geometry-aware algorithm.

\subsubsection{Feasible Region for $\boldsymbol{p}_b$ Convexification}
The non-convex constraint (\ref{BS limits}) for the feasible region of $\boldsymbol{p}_b$ needs to be transformed into a convex one by defining a larger region $C_3$, which fully encompasses $C_1$ and has a convex shape in Fig.~\ref{bound_def}, as the new boundary for $\boldsymbol{p}_b$. Then, the objective function in (\ref{P0}) is reformulated as $J_{\text{p}}(\boldsymbol{\zeta})$ with penalty
\begin{align}
    \tag{P1}
    \operatorname*{argmin}_{\boldsymbol{\zeta}} 
      J_{\text{p}}(\boldsymbol{\zeta}) = J(\boldsymbol{\zeta})
    + \tau_{\text{p}} \delta \cdot d(\boldsymbol{p}_b, C_1) 
    \label{p1}
\end{align}
subject to:
\begin{align}
\delta &=
\begin{cases}
1, & \text{if } \boldsymbol{p}_b \in \mathcal{C}_3 \setminus \mathcal{C}_1,\\
0, & \text{otherwise},
\end{cases}\label{eq:delta_indi}\\[3pt]
& \boldsymbol{p}_b \in C_3, (\ref{cov_c})\ \text{and}\ (\ref{EIRP}), \notag
\end{align}
where $d(\boldsymbol{p}_b, C_1)$ is part of the penalty function representing the smallest distance between $\boldsymbol{p}_b$ and boundary $C_1$. The term $\delta$ serves as a binary indicator with $\delta = 1$ if $\boldsymbol{p}_b$ is within the region that lies inside $C_3$ but outside $C_1$ as stated in $C_3 \setminus C_1$, otherwise it equals 0 as defined in (\ref{eq:delta_indi}). The penalty coefficient is $\tau_{\text{p}}$, which needs to be a sufficiently large finite value to draw the $\boldsymbol{p}_b$ towards the interior of the \(C_1\) contour. Provided $\tau_{\text{p}}$ is sufficiently large, it does not directly affect the optimization results but merely restricts the feasible region.

\subsubsection{The \ac{NM} Method for EMF Exposure Minimization}
\label{sec:nm} 
This section reverts to using $\boldsymbol{p}_b,EIRP$ instead of  $\boldsymbol{\zeta}$, since the optimization algorithm now treats these two parameters explicitly.
Given the aforementioned properties of the problem, the \ac{NM} simplex method is selected as the suitable solution for several reasons. First, it is derivative-free and therefore appropriate for black-box optimization problem, where only objective values $J_{\text{p}}(\boldsymbol{p}_b,EIRP)$ are available. Second, despite the significant reduction in computational complexity achieved by Algo.~\ref{algo_2}, hundreds of \ac{RL} simulations are still required for estimating the \ac{EMF} exposure over the target area. The \ac{NM} method is therefore preferable since it is recognized as a fast-converging method \cite{wright1998optimization}, \cite{liu2024enhanced}. Finally, the \ac{NM} method adapts its simplex geometrically to local curvature, which is well-suited to geometry-dependent optimization. With this choice, problem (\ref{p1}) is solved by the EMF exposure minimization algorithm that iteratively performs two steps: first, the minimum $EIRP$ is determined for a specific position of the BS, and then the BS position set is updated via the \ac{NM} method based on the derived $J_{\text{p}}(\boldsymbol{p}_b,EIRP)$. Following each BS position update, all stages illustrated in Fig.~\ref{fig:workflow} are repeated until the convergence criterion is satisfied, yielding the optimal $\boldsymbol{p}_{b,\text{op}}$ and $EIRP_{\text{op}}$. This finalizes the workflow for joint optimization of BS deployment and power based on deterministic channel simulations.

\begin{algorithm}
 \caption{\ac{EMF} exposure minimization algorithm}
 \begin{algorithmic}[1]
 \renewcommand{\algorithmicrequire}{\textbf{Input:}}
 \renewcommand{\algorithmicensure}{\textbf{Output:}}
 \REQUIRE $N_s$
 \STATE \textbf{Initialization}: Random BS position set $\mathcal{B}^0$ with $q = 0$
   \STATE Obtain $J_{\text{p},n_s}^{q}$ for $n_s = \{1,2,...,N_s+1\}$, forming $\mathcal{J}_{\text{p}}^q$
    \STATE Ordering$\mathcal{J}_{\text{p}}^q$ into $\tilde{\mathcal{J}}_{\text{p}}^q$ and corresponding $\mathcal{B}^q$ into $\tilde{\mathcal{B}}^q$
    \WHILE{ $\max\mathcal{D}_{q} > D_{\text{max}}$}
    \STATE Reflection: calculate $\boldsymbol{p}_{b,r}^q$ with $\mu = \mu_r$ using (\ref{p_miu_equation}) and obtain $J_{\text{p},r}^q$
    \IF {$\tilde{J}_{\text{p},1}^q \leq J_{\text{p},r}^q < \tilde{J}_{\text{p},N_s}^q$}
        \STATE $\boldsymbol{p}_{b,N_s+1}^{q+1} = \boldsymbol{p}_{b,r}^q$ and go to line 31
    \ELSIF {$J_{\text{p},r}^q < \tilde{J}_{\text{p},1}^q$}
        \STATE Expansion: calculate $\boldsymbol{p}_{b,e}^q$ with $\mu = \mu_e$ using (\ref{p_miu_equation}) and obtain $J_{\text{p},e}^q$
        \IF {$J_{\text{p},e}^q < J_{\text{p},r}^q$}
            \STATE $\boldsymbol{p}_{b,N_s+1}^{q+1}  = \boldsymbol{p}_{b,e}^q$ and go to line 31
        \ELSE
            \STATE $\boldsymbol{p}_{b,N_s+1}^{q+1}  = \boldsymbol{p}_{b,r}^q$ and go to line 31
        \ENDIF
    \ELSIF {$\tilde{J}_{\text{p},N_s}^q< J_{\text{p},r}^q < \tilde{J}_{\text{p},N_s+1}^q$}
        \STATE Outer contraction: calculate $\boldsymbol{p}_{b,oc}^q$ with $\mu = \mu_{oc}$ using (\ref{p_miu_equation}) and obtain $J_{\text{p},oc}^q$
        \IF {$J_{\text{p},oc}^q < J_{\text{p},r}^q$}
            \STATE $\boldsymbol{p}_{b,N_s+1}^{q+1}  = \boldsymbol{p}_{b,oc}^q$ and go to line 31
        \ELSE
            \STATE  go to line 30
        \ENDIF
    \ELSIF {$\tilde{J}_{\text{p},N_s+1}^q \leq J_{\text{p},r}^q$}
        \STATE Inner contraction: calculate $\boldsymbol{p}_{b,ic}^q$ with $\mu=\mu_{ic}$ using (\ref{p_miu_equation}) and obtain $J_{\text{p},ic}^q$
        \IF {$J_{\text{p},ic}^q < \tilde{J}_{\text{p},N_s+1}^q$}
            \STATE $\boldsymbol{p}_{b,N_s+1}^{q+1}  = \boldsymbol{p}_{b,ic}^q$ and go to line 31
        \ELSE
            \STATE go to line 30
        \ENDIF
    \ENDIF
    \STATE Shrink: formulate $\mathcal{B}^{q+1}$ with $\tilde{\boldsymbol{p}}_{b,n_s}^{q+1} \!=\! \tilde{\boldsymbol{p}}_{b,1}^q \!+\! \mu_s (\tilde{\boldsymbol{p}}_{b,n_s}^q \!-\! \tilde{\boldsymbol{p}}_{b,1}^q)$ for $2\!\leq\! n_s\!\leq\! N_s+1$, $q = q+1$ and go to line 2
    \STATE Formulate $\mathcal{B}^{q+1} = \{ \tilde{\boldsymbol{p}}_{b,1}^q,...,  \tilde{\boldsymbol{p}}_{b,n_s}^q,..., \boldsymbol{p}_{b,N_s+1}^q \}$, $q = q+1$ and go to line 2
    \ENDWHILE
 \ENSURE $\boldsymbol{p}_{b,\text{op}}=\tilde{\boldsymbol{p}}_{b,1}^q$, $EIRP_{op} = \tilde{EIRP}_{1}^q$
 \end{algorithmic} 
 \label{algo3}
\end{algorithm}

The overall algorithm is summarized in Algo.~\ref{algo3}. The \ac{NM} method operates on an initial position set $\mathcal{B}^0 = \{\boldsymbol{p}_{b,1}^0,...,\boldsymbol{p}_{b,n_s}^0, ..., \boldsymbol{p}_{b,N_s+1}^0\}$ with $n_s\in\{1,2,...,N_s+1\}$, where each element represents a random BS position $\boldsymbol{p}_{b}$. 
In each $q$-th iteration, for a specific $\boldsymbol{p}_{b,n_s}^q$, $EIRP_{n_s}^q$ in (\ref{p1}) is first determined. 
Under the presence of intercell interference, $EIRP_{n_s}^q$ is determined by the minimum \ac{EIRP} required to satisfy the coverage constraint (\ref{cov_c}) using \eqref{sinr_ideal}. Since EMF exposure increases monotonically with EIRP for a specific BS position, as described in Section~\ref{BS_place_problem_and_NM}, selecting the minimum EIRP satisfying the coverage constraint minimizes the EMF exposure objective in \eqref{P0}, which decouples power determination from position optimization, reducing the search dimensionality.

With known $EIRP_{n_s}^q$, each function value $J_{\text{p},n_s}^{q} = J_{\text{p}}(\boldsymbol{p}_{b,n_s}^q, EIRP_{n_s}^q)$ is obtained, forming the function value set $\mathcal{J}_{\text{p}}^q$ in line 2. Throughout, the indexing is kept consistent so that whenever $\boldsymbol{p}_b$ carries subscripts or superscripts, the corresponding $EIRP$ and $J_{\text{p}}$ adopt the same indices, indicating these parameters are determined for the specific position. 
Then, $\mathcal{J}_{\text{p}}^q$ is ordered ascendingly into $\tilde{\mathcal{J}}_{\text{p}}^q = [\tilde{J}_{\text{p},1}^{q},...,\tilde{J}_{\text{p},n_s}^{q},...,\tilde{J}_{\text{p},N_s+1}^{q}]$ and the correspondingly ordered position set is denoted as $\tilde{\mathcal{B}}^q = \{\tilde{\boldsymbol{p}}_{b,1}^{q},...,\tilde{\boldsymbol{p}}_{b,n_s}^q, ..., \tilde{\boldsymbol{p}}_{b,N_s+1}^q\}$ in line 3, where $\tilde{\boldsymbol{p}}_{b,1}^q$ and $\tilde{\boldsymbol{p}}_{b,N_s+1}^q$ are defined as the best and worst positions. Afterwards, the algorithm attempts to update $\tilde{\boldsymbol{p}}_{b,N_s+1}^q$ with a new position $\boldsymbol{p}_{b,N_s+1}^{q+1}$ of the form
\begin{equation} 
     \boldsymbol{p}_{b,N_s+1}^{q+1} = (1+\mu)\bar{\boldsymbol{p}^{q}} - \mu \tilde{\boldsymbol{p}}_{b,N_s+1}^q,
     \label{p_miu_equation}
\end{equation}
where $\bar{\boldsymbol{p}^q}= \frac{1}{N_s} \sum_{n_s=1}^{N_s} \tilde{\boldsymbol{p}}_{b,n_s}^q$ is the centroid of the $\tilde{\mathcal{B}}^q$ excluding the worst position. By adjusting the value of $\mu$, $\boldsymbol{p}_{b,N_s+1}^{q+1}$ varies accordingly and is indexed based on the specific conditions encountered. The Algo.~\ref{algo3} includes two main transformations. Reflection in line 5 moves away from the worst position, and shrink in line 30 contracts toward the best position. The universal choices for $\mu$ in the standard NM algorithm are: reflection $\mu_r= 1$, expansion $\mu_e= 2$, outer contraction $\mu_{oc}= 0.5$, inner contraction $\mu_{ic}= -0.5$, and shrink $\mu_s = 0.5$. 
The convergence criterion of the NM method is based on the distance between each position in $\mathcal{B}^{q}$ with
\begin{equation} \label{NM_distance}
    \mathcal{D}_{q} = \left\{\! \left\| \boldsymbol{p}_{b,i}^q - \boldsymbol{p}_{b,j}^q \right\| \;\middle| i,j \!\in\! \{1,2, \ldots, N_s+1\}, i \neq j \right\},
\end{equation}
where $\mathcal{D}_{q}$ includes all the pair-wise distance between any two positions in the $q$-th iteration. The Algo.~\ref{algo3} converges when all the positions in $\mathcal{B}^{q}$ are close enough satisfying $\max\mathcal{D}_{q} \leq D_{\text{max}}$ in line 4 with $D_{\text{max}}$ the maximum distance. Distance is used as the convergence criterion because practical BS installation admits positional tolerance. Finally, $\boldsymbol{p}_{b,\text{op}}$ is derived as the best position after ordering and the corresponding $EIRP_{\text{op}}$ as the output of Algo.~\ref{algo3}.
In general, by determining the minimum transmit power required to satisfy the coverage constraint at each BS position and using the corresponding EMF exposure to guide the NM position updates, the algorithm enables joint optimization of BS deployment and power from deterministic channel simulations despite the complex spatial variation of the objective and the lack of its analytical gradients.

\section{Results Analysis}
\label{sec:experiment_results}
In this section, relevant simulation settings are introduced. The simulation results of Algo.~\ref{algo_2}, optimized results for (\ref{p1}), and the comparison results with the baseline are presented and analyzed.

\subsection{Simulation Settings}
\begin{table}
\begin{center}
\caption{Simulation settings overview}
\label{sim_set}
\begin{tabular}{| c | c | c | c| }
\hline
Parameters & value & Parameters & value \\
\hline
$z_b$ & \SI{30}{m}  &  $\Delta P_{R,\text{max}}$ &\SI{5}{dB} \\
\hline
$z_n$ & \SI{1.5}{m} &$g_0$ &\SI{1.25}{m} \\
\hline
$f_0$  & \SI{3.5}{GHz} & $g_1$ &\SI{10}{m}\\ 
\hline
$M_p$ & 5 & $T_{\text{avg}}$ &\SI{6}{min}\\ 
\hline
$\sigma_{n}^2$  & \SI{-95}{dBm} &  $\delta_{\text{tol}}$ & \SI{0.005}{V/m}\\ 
\hline 
$\boldsymbol{p}_{l},l=1$  & [-146, -36, 30] & $\boldsymbol{p}_{l},l=2$ & [290, 20, 30]\\ 
\hline
$P_{T,l}$  & \SI{200}{W} & $N_s$  & 7 \\ 
\hline
$G_{T,l}$  & \SI{15}{dBi} & $\rho_{\text{entry}}$ & 0.1 \\ 
\hline
$\rho_{\text{cov}}$ & 0.99 & $\tau_{e}^2$& 0.1 \\
\hline
$EIRP_{max}$  & \SI{81.18}{dBm} &$D_{\text{max}}$ &\SI{1}{m}\\ 
\hline 
$\gamma_{\text{min}}$  & \SI{5}{dB} & $\upsilon$ & \SI{5}{km/h}\\ 
\hline
\end{tabular}
\end{center}
\end{table}
Simulation parameters used are listed in Table~\ref{sim_set}, with some of them described below. The power of BS transmitting in the 3450-\SI{3550}{MHz} (n77) band is limited to an \ac{EIRP} of \SI{1640}{Watts/MHz} according to \ac{FCC} \cite{act2015before}. Considering the total bandwidth of \SI{80}{MHz} \cite{10500988}, the maximal $EIRP_{\text{max}}$ is \SI{81.18}{dBm} and the noise power is \SI{-95}{dBm}. For beamforming, the BS is configured to cover the 5 strongest multipaths with $M_p=5$. In Algo.~\ref{algo_2}, $\Delta P_{R,\text{max}}$ must be specified to indicate significant changes in channel conditions. In the service mode of mobile communications, the \ac{RSRP} is categorized into four different groups \cite{putra20214g}: excellent ($\geq$ \SI{-80}{dBm}), good (\SI{-90}{dBm} to \SI{-80}{dBm}), fair to poor (\SI{-100}{dBm} to \SI{-90}{dBm}), and no signal. Accordingly, $\Delta P_{R,\text{max}} =\SI{5}{dB}$ is adopted so that if the $P_{R,\text{bf}}$ difference between neighboring positions exceeds \SI{5}{dB}, the intermediate point is likely to fall into a different \ac{RSRP} level. 
When applying \ac{NM} method, the initial 8 positions in $\mathcal{B}^0$ are selected as in Fig.~\ref{bound_def} and the algorithm will converge with $D_{\text{max}} = \SI{1}{m}$. The UE number dependent $P_{\text{RF}}$ values are taken from \cite{10500988} and over $N\in[18,45]$, $P_{\text{RF}}$ range from 8.3\% to 16.3\%. Consequently, $N(\omega)$ in (\ref{MC avg}) is also constrained to this interval. The interfering BS configurations follow \cite{chiaraviglio2021health} with maximum transmit power of \SI{200}{W} and antenna gain of \SI{15}{dBi}. The \ac{CSI} error is set to $\tau_e^2=0.1$ \cite{mi2017massive}.

\subsection{\ac{AGR} Results for \ac{EMF} Exposure Estimation}
\begin{figure}[!t]
\centering
\subfigure[]{\includegraphics[width=1.6in, trim=0mm 5mm 0mm 0mm,clip]{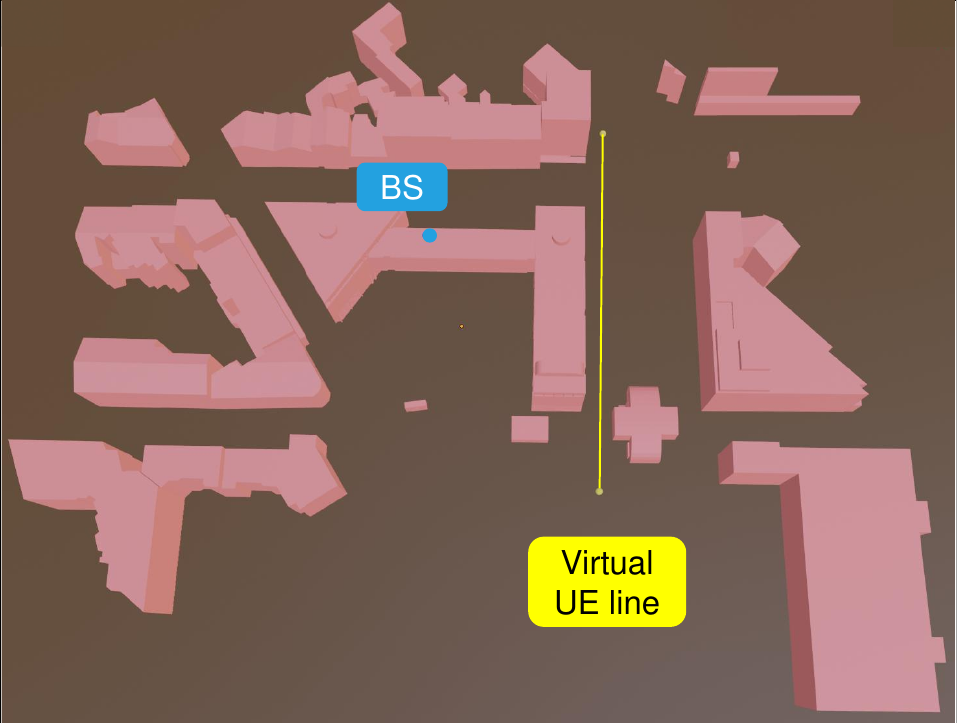}%
\label{sim_scene_2}}
\hspace{-5pt}
\subfigure[]{\includegraphics[width=1.6in]{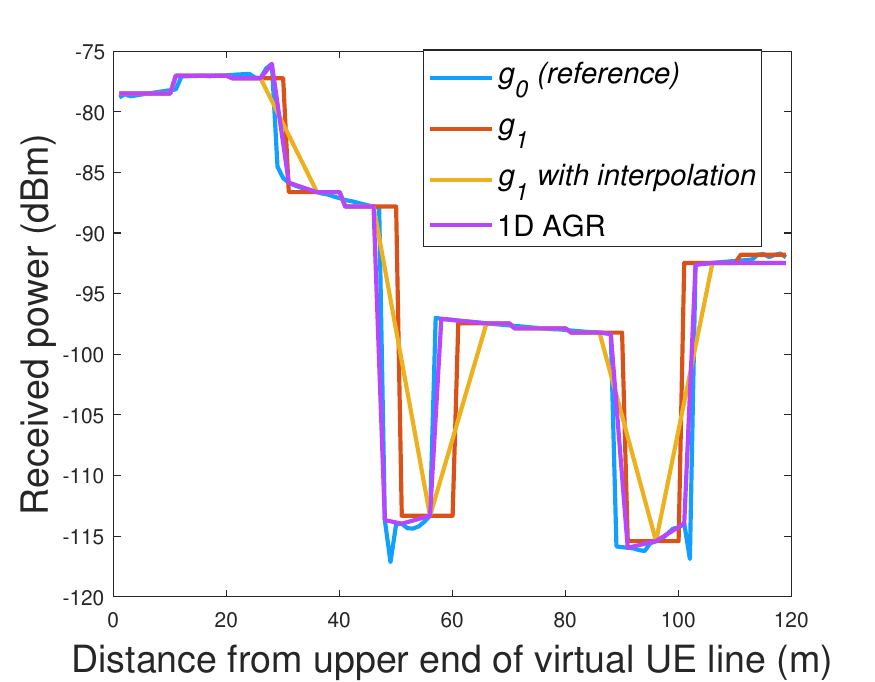}%
\label{1d_results}}
\caption{(a) 3D scenario as illustration of \ac{AGR} algorithm in 1D case. The BS is at \raisebox{-0.5ex}{\color[rgb]{0.55,0.95,1}\Large \textbullet} with unit \ac{EIRP} and the virtual \ac{UE} follows \raisebox{0.5ex}{\textcolor{yellow!75!orange}{\rule{1.4em}{1.2pt}}}. (b) Comparison for different scheme of \ac{EMF} exposure estimation to $g_0$ of 1D case. }
\end{figure}

The purpose of \ac{AGR} can be illustrated from a numerical analysis perspective, a 1D case is demonstrated by applying different grid lengths to predict \ac{EMF} exposure. 
Fig.~\ref{sim_scene_2} illustrates the simulation scenario, in which the virtual \ac{UE}s are aligned in a straight line, which is uniformly discretized with a reference grid length $g_0$, resulting in \ac{EMF} exposure shown as the blue curve in Fig.~\ref{1d_results}. However, this exhaustive simulation might be excessively time-consuming for an urban area and is inefficient, as there are regions where the large-scale fading characteristics are similar. To address this issue, a larger grid length $g_1$ is chosen and its \ac{EMF} exposure prediction is depicted by the red line in Fig.~\ref{1d_results}. 
It is obvious that such a large grid cannot represent the channel accurately compared with $g_0$, resulting in a \ac{std} $\sigma$ of \SI{38}{dB} between the two, as shown in Table \ref{tab1}. Significant errors occur due to the failure to predict sudden changes such as around 47 or \SI{56}{m}. As a comparison, directly interpolating the results of $g_1$, as shown by the yellow curve in Fig.~\ref{1d_results}, reduces $\sigma$ but increases the mean of difference $\eta$. As a result, the \ac{AGR} algorithm can refine the grid to better capture the changing characteristic of the channel, leading to a flexible grid length, with the results shown as the purple line in Fig.~\ref{1d_results} and the $\eta$ and $\sigma$ are significantly reduced, as indicated in Table \ref{tab1}.
\begin{table}
\begin{center}
\caption{Results compared to reference grid length $g_0$.}
\label{tab1}
\begin{tabular}{| c | c | c |}
\hline
Method & $\eta$ (dB) & $\sigma$ (dB)\\
\hline
Grid length $g_1$  & 2.28 &  38.06  \\
\hline
Grid length $g_1$ with interpolation & 2.79 & 21.50\\
\hline
1D \ac{AGR} & 0.68 & 4.99\\ 
\hline 
\end{tabular}
\end{center}
\end{table}

\begin{figure*}[!t]
\centering
\subfigure[]{\includegraphics[width=1.82in,trim=0.7mm 0.22mm 2mm 7.5mm,clip]{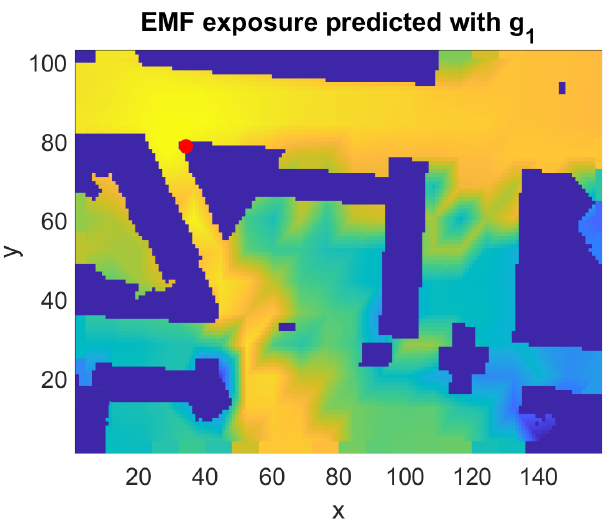}%
\label{10m_2d_1}}
\hfil
\subfigure[]{\raisebox{1pt}{\includegraphics[width=1.8in,trim=0.6mm 0.3mm 2mm 4.5mm,clip]{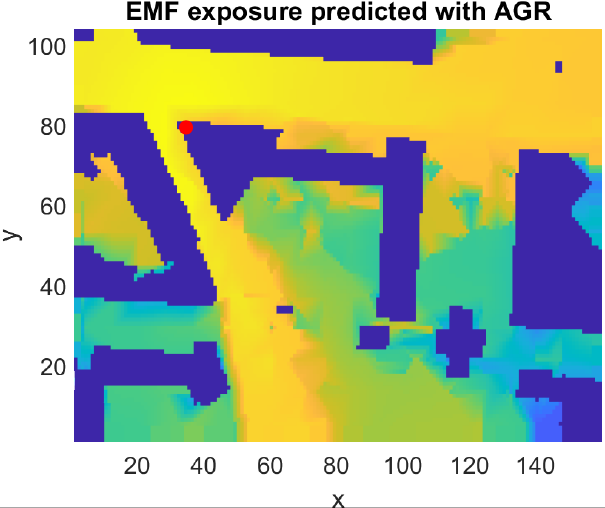}}%
\label{10m_2d_hdii}}
\hfil
\subfigure[]{\raisebox{0pt}{\includegraphics[width=2.0in,trim=0.7mm 0.32mm 1mm 7.4mm,clip]{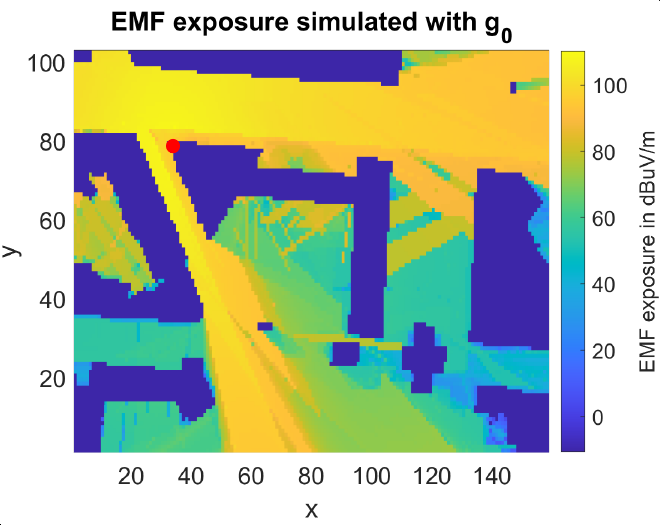}}%
\label{125m_fine}}
\caption{(a) Simulated \ac{EMF} exposure with $g_1$ and afterwards directly performs a 2D interpolation. (b) Simulated \ac{EMF} exposure using Algo.~\ref{algo_2}. (c) Simulated with finest $g_0$ as the reference simulation. All three subfigures share one colorbar and the BS positions at \raisebox{-0.5ex}{\color[rgb]{1,0,0}\Large \textbullet} with unit \ac{EIRP}. The area shown corresponds to the \ac{EMF} exposure within the boundary $C_2$ depicted in Fig.~\ref{bound_def}, while the blue areas represent buildings. The $x$ and $y$ axes represent coordinates along the west-east and north-south direction, respectively. Each grid in the subfigures corresponds to a distance of \SI{1.25}{m}, which is the minimum grid length $g_0$ along both the $x$ and $y$ axes.}

\end{figure*}

Subsequently, a 2D prediction example is demonstrated in the scenario of Fig.~\ref{bound_def} to demonstrate the efficiency of the \ac{AGR} algorithm. Fig.~\ref{10m_2d_1} shows the initial results in line 3 in Algo.~\ref{algo_2} with $g_1$, and a 2D interpolation is directly performed afterwards, requiring 260 simulations. This can only reflect the general \ac{EMF} exposure distribution compared to the results using $g_0$, which serves as the reference as in Fig.~\ref{125m_fine}, requiring in a total of \(\mathcal{U}_0\) 16,365 simulations. After performing the \ac{AGR} algorithm with \(\mathcal{U}_{\text{tot}}\) 665 simulations, the \ac{EMF} exposure is more finely represented, as shown in Fig.~\ref{10m_2d_hdii}. Only 4.0\% of the simulations were performed compared with $g_0$, yet $\eta$ is \SI{2.74}{dB}, and $\sigma$ is \SI{5.33}{dB}. Other uniform grid lengths are also simulated to compare and are shown in Table \ref{table_compare_HDII}. The results of Algo.~\ref{algo_2} demonstrate comparable accuracy while requiring less than half of the simulations compared to using the grid length of \SI{3.75}{m}. Considering the trade-off between accuracy and computational effort, \ac{AGR} offers a flexible approach to refining grid when necessary, accurately predicting overall \ac{EMF} exposure while significantly reducing the computation complexity.
\begin{table}
\begin{center}
\caption{The accuracy for different simulated grid length to reference $g_0$.}
\label{table_compare_HDII}
\begin{tabular}{| c | c | c | c |}
\hline
Grid length (m) &  $\eta$ (dB) & $\sigma$ (dB) & No. of simulations\\
\hline
10 & 4.92 & 7.74 & 260\\
\hline
7.5 & 4.15 & 6.22 & 484\\
\hline 
5 & 3.21 & 5.49 & 1,040\\ 
\hline 
3.75 & 2.63 & 5.09 & 1,851\\ 
\hline 
2.5 & 2.24 & 4.63 & 4,154\\ 
\hline 
Reference: 1.25 & 0 & 0 & 16,365\\ 
\hline 
Algo.~\ref{algo_2} & 2.74 & 5.33 & 665\\ 
\hline 
\end{tabular}
\end{center}
\label{pl_para}
\end{table}

\begin{figure*}[!t]
\centering
\subfigure[]{\includegraphics[width=2.46in, trim=5mm 2mm 5mm 2mm,clip]{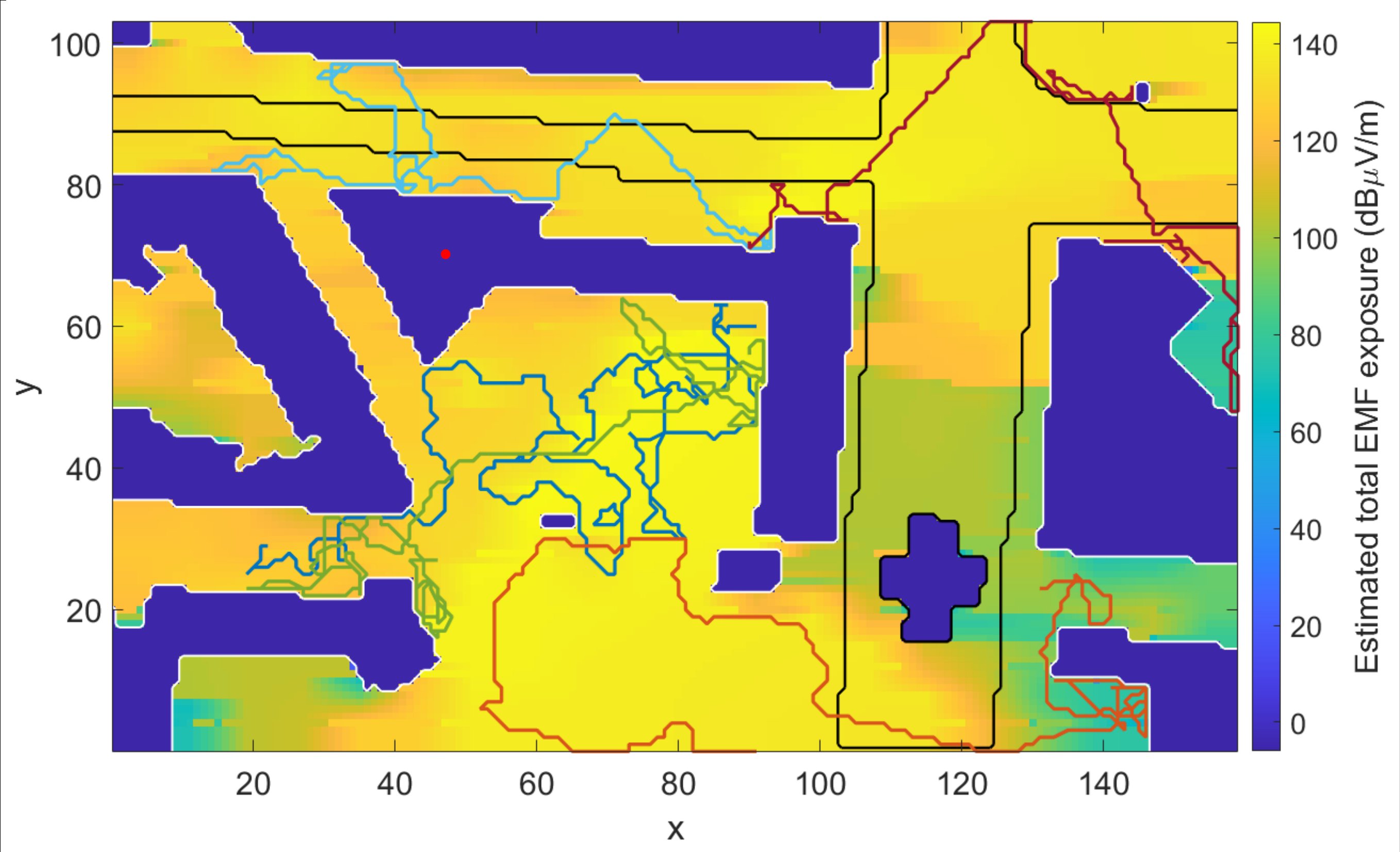}%
\label{P2_opt}}
\subfigure[]{\includegraphics[width=2.40in]{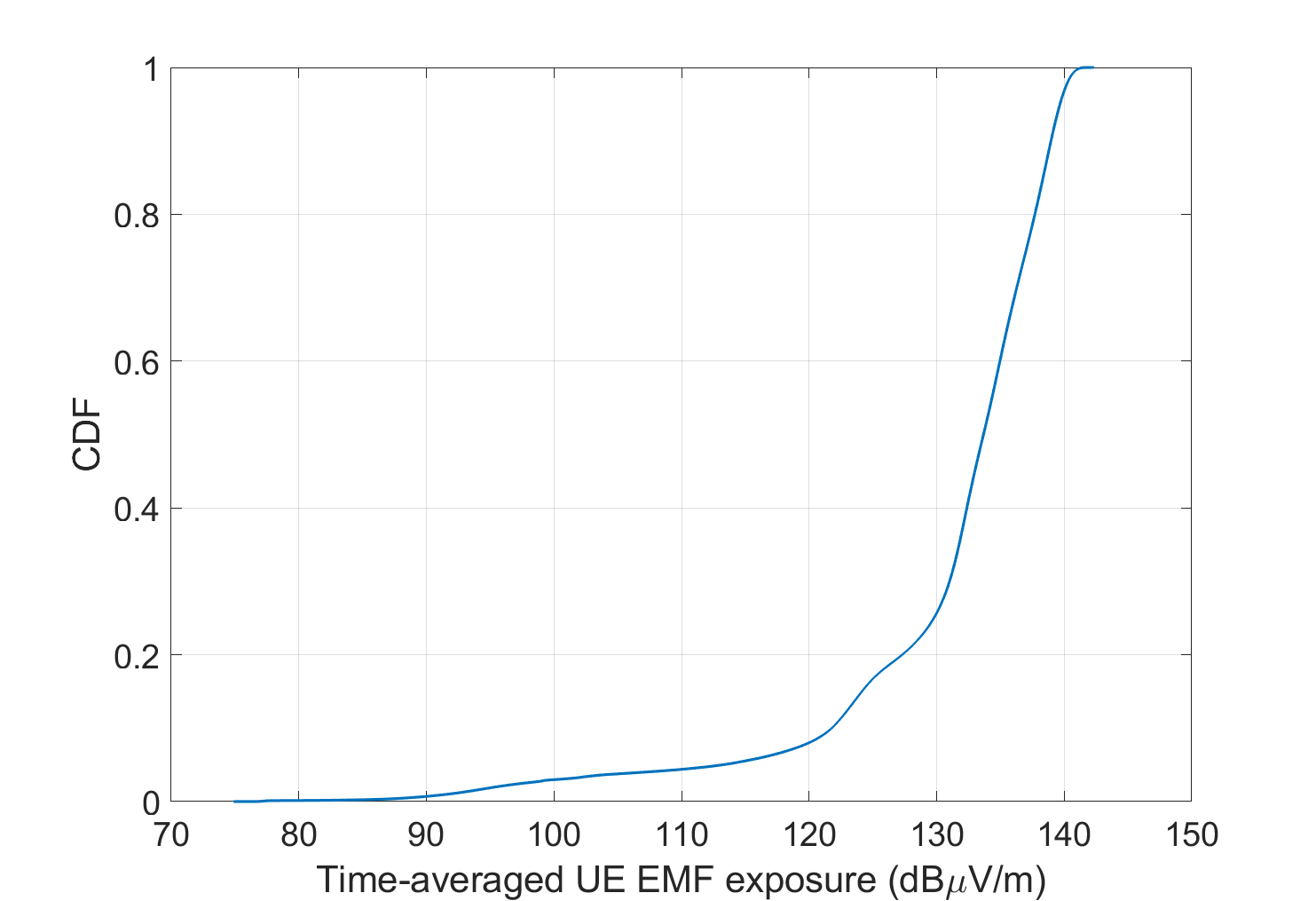}%
\hspace{-5pt}
\label{CDF}}
\subfigure[]{\raisebox{-4pt}{\includegraphics[width=2.15in]{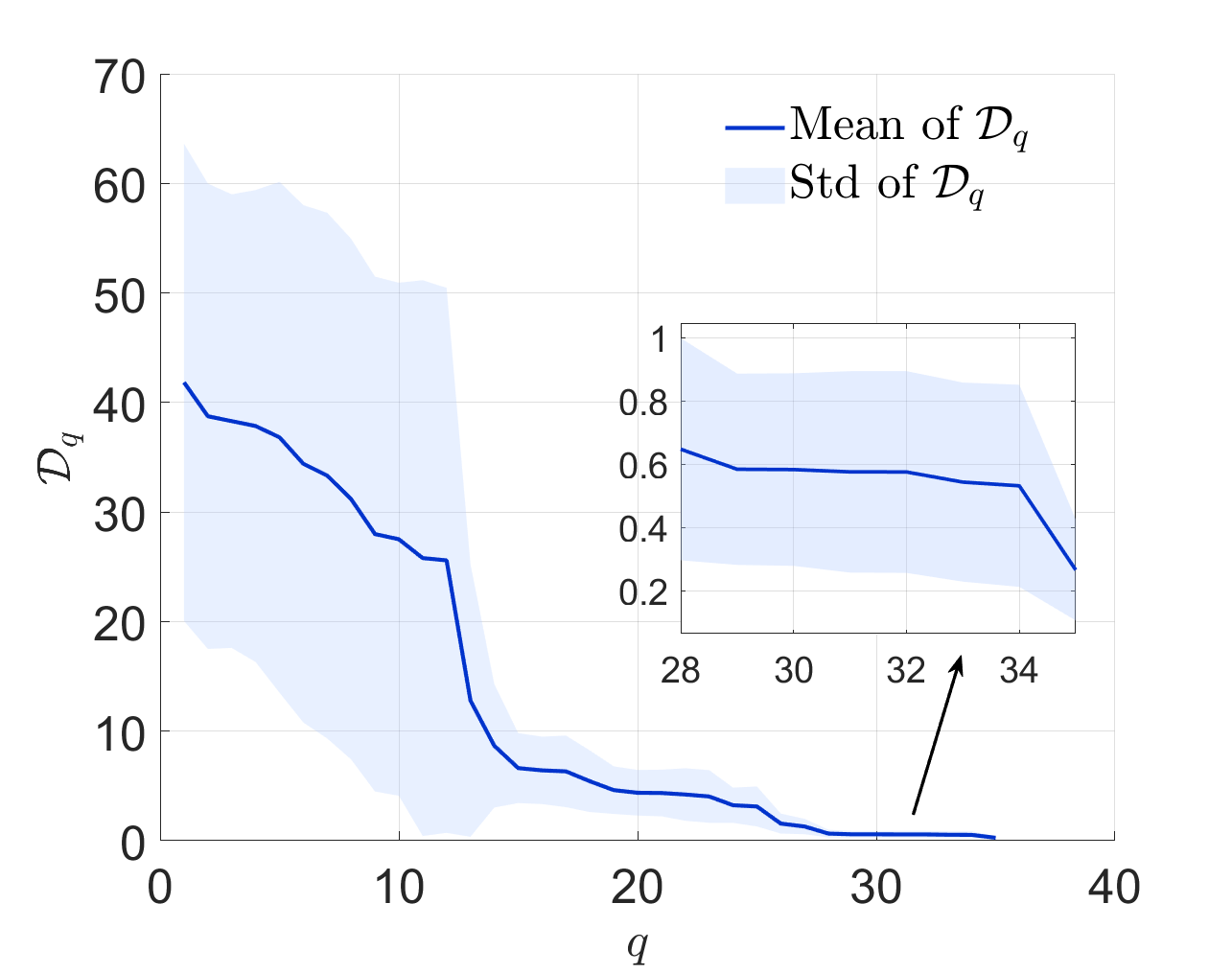}}%
\hspace{-5pt}
\label{NM_conv}}
\caption{(a) The estimated \ac{EMF} exposure over target area with $b_{op}$ indicated by \raisebox{-0.5ex}{\color[rgb]{1,0,0}\Large \textbullet} and $EIRP_{op}$ of \SI{72.43}{dBm} for (\ref{p1}). The black contour outlines the street area, and five example \ac{UE} trajectories are illustrated. (b) \ac{CDF} of time-averaged \ac{EMF} exposure for \ac{UE} with $\boldsymbol{p}_{b,\text{op}},EIRP_{\text{op}}$. (c) Convergence behavior for Algo.~\ref{algo3}.}
\vspace{-1\baselineskip}
\end{figure*}

\subsection{Joint Optimization Results for P1}
\label{joint_results}
Fig.~\ref{P2_opt} depicts the optimized BS deployment $\boldsymbol{p}_{b,\text{op}} = [14.05,32.34,30]$ for problem (\ref{p1}), and the estimated \ac{EMF} exposure with $EIRP_{\text{op}}=\SI{72.43}{dBm}$, which includes the effect from two interfering BSs and attains coverage of 99\%. Five examples of the \ac{UE} trajectories generated according to Section~\ref{UE mobility} are illustrated in Fig.~\ref{P2_opt}, demonstrating different tendencies to walk straight, consistent heading at street crossings and window-shopping behavior. The \ac{CDF} of the time-averaged \ac{EMF} exposure for the dynamic \ac{UE} under optimized deployment is shown in Fig.~\ref{CDF}, including $\omega=40,769$ \ac{MC} simulations. It shows a concentrated distribution of \ac{EMF} exposure with mean \SI{133.96}{\dB\micro\volt/\meter} and 95th percentile of \SI{139.67}{\dB\micro\volt/\meter}, which is still \SI{16.04}{dB} below the regulation limit, providing a sufficient compliance margin. The convergence behavior of Algo.~\ref{algo3} is presented in Fig.~\ref{NM_conv}. The mean and \ac{std} of pairwise distance set $\mathcal{D}_{q}$ in (\ref{NM_distance}) are initially large, indicating that $\mathcal{B}^0$ spans the deployable region sufficiently for global exploration, and decrease steadily in general. The sharp drops at 12, 25 and 34 iterations correspond to the shrink process in Algo.~\ref{algo3}, where all the positions in $\tilde{\mathcal{B}}^q$ are drawn towards the best position. After 35 iterations, the convergence condition is satisfied, where all positions in $\mathcal{B}^q$ lie within the sphere of diameter $D_{\max}$, making further improvement less probable, as highlighted in the zoom-in plot of Fig.~\ref{NM_conv}. 

\subsection{Comparison to Empirical Channel Model}
As mentioned in Section \ref{introduction}, most BS deployment optimizations are formulated using empirical channel models. Thus, a comparison with the solution using an empirical channel model is meaningful to demonstrate the benefits of the proposed workflow. To ensure optimal performance, an empirical channel model tailored to the target area is utilized. Section \ref{validation} has demonstrated that Algo.~\ref{alg:Simplified_Hybrid_SBR_RT} is sufficiently accurate to be comparable with real-world measurements. Therefore, based on the received power simulation results in line 3 of Algo.~\ref{algo_2} with grid length $g_1$, a \ac{CI} channel model \cite{zhang2019measurement} is derived for path loss
\begin{equation}
PL(f_0, d_{\text{3D}}) [\text{dB}] = PL(f_0, d_0) + 10n_{\text{PLE}} \log_{10}\left(d_{\text{3D}}/{d_0}\right) + X
\label{PL_equa}
\end{equation}
where $d_{\text{3D}} = ||\boldsymbol{p}_b-\boldsymbol{u}_{(i,j)}^1||$ with $\boldsymbol{u}_{(i,j)}^1\in\mathcal{U}_1$. The variable $n_{\text{PLE}}$ represents the path loss exponent and $X$ denotes the shadowing factor, which accounts for the large-scale signal fluctuations due to obstacles and environmental factors, and is usually a zero-mean Gaussian distributed random variable. The term $PL(f_0, d_0)$ is the reference path loss at the distance $d_0$, and in \ac{LoS} condition $PL(f_0, d_0) [\text{dB}] = 20 \log_{10} \left(4 \pi d_0/\lambda\right)$.
Whereas in \ac{NLoS} condition, one of the simulated path loss will be used as the reference.

\begin{table}[h]
\centering
\caption{Parameters of the CI model}
\label{pl_para}
\begin{tabular}{|c|c|c|c|c|}
\hline
\multirow{2}{*}{Parameters} & \multicolumn{2}{c|}{Sim. results} & \multicolumn{2}{c|}{Results from \cite{zhang2019measurement}} \\ \cline{2-5} 
                  & \ac{LoS} & \ac{NLoS}     & \ac{LoS} & \ac{NLoS}    \\ \hline
$n_{\text{PLE}}$            & 2.00 &  2.58          & 2.15 & 2.46          \\ \hline
$\sigma$           & 0.67&	7.57          & 3.26     & 6.17           \\ \hline
\end{tabular}
\end{table}
All the simulated $PL$ are fitted using equation (\ref{PL_equa}), and the resulting parameters are summarized in Table~\ref{pl_para}. Other than $n_{\text{PLE}}$, \ac{std} of the shadowing factor $\sigma$ quantifies the variability in the received power due to shadowing effects. In simulated results from this paper, $n_{\text{PLE}}$ is 2.00 under \ac{LoS} conditions, indicating free-space propagation. For \ac{NLoS} conditions, $n_{\text{PLE}}$ increases to 2.58 due to additional attenuation from buildings. These values are comparable to urban macro measurements at \SI{3.5}{GHz} from \cite{zhang2019measurement} in the right column of Table~\ref{pl_para}.
The $n_{\text{PLE}}$ and $\sigma$ for the \ac{LoS} case in this paper appear relatively more accurate, as the simulation scenario is more stable than real-world conditions and free from influences of measurement equipment. For the \ac{NLoS} case, the results are also comparable, though with a slightly higher $\sigma$, likely due to the simulated position of \ac{Tx} being at the center of the scenario, resulting in a more diverse \ac{NLoS} channel compared to \cite{zhang2019measurement}, where the \ac{Tx} is positioned at the edge of the scenario. Such agreement in Table~\ref{pl_para} supports the validity of our \ac{RL} simulation as well. In many research such as \cite{al2014optimal}, channel is often modeled probabilistically with $P_{\text{LoS}}$ and $P_{\text{NLoS}}$ representing the probability of \ac{LoS} and \ac{NLoS} channel due to the lack of information about the exact type of channel present. And the \ac{LoS} probability in urban macro scenario is \cite{al2014optimal}:
\begin{equation}
P_{\text{LoS}} = 
\begin{cases}
\begin{aligned}
& 1, & \quad d_{\text{2D}} \leq 18 \\[8pt]
& 18/d_{\text{2D}} \!+\! \exp\left(-d_{\text{2D}}/63\right)\left(1 \!-\! 18/d_{\text{2D}}\right), & \quad d_{\text{2D}} > 18
\end{aligned}
\end{cases}
\label{possibility}
\end{equation}
where $d_{\text{2D}}$ is the Euclidean distance in the horizontal plane between \ac{Tx} and \ac{Rx} and $P_{\text{NLoS}} = 1- P_{\text{LoS}}$. A random variable is generated to compare with the distance-dependent \ac{LoS} probability and if it is less than the \ac{LoS} probability, the channel is \ac{LoS}, and otherwise \ac{NLoS}. Utilizing the empirical channel model in (\ref{PL_equa}), the \ac{EMF} exposure is estimated in an open area. 
Solving (\ref{p1}) yields the optimized BS position $\boldsymbol{p}_{b,\text{op}}^{\text{e}} = [87.25,37.62,30]$ and $EIRP_{\text{op}}^{\text{e}} = \SI{64.49}{dBm}$, satisfying the coverage requirement with an average \ac{UE} \ac{EMF} exposure of \SI{130.94}{\dB\micro\volt/\meter}. The optimized BS position $\boldsymbol{p}_{b,\text{op}}^{\text{e}}$ differs substantially from $\boldsymbol{p}_{b,\text{op}}$ in Fig.~\ref{P2_opt}, with a separation of \SI{73.48}{m}, and yields an average \ac{EMF} exposure \SI{3.02}{dB} lower than that shown in Fig.~\ref{CDF}. 
The empirical channel model predicts a lower EIRP requirement for the target coverage, resulting in an underestimation of EMF exposure, potentially increasing the risk of exceeding regulatory limits. When the scenario geometry is explicitly accounted for, the BS at $\boldsymbol{p}_{b,\text{op}}^{\text{e}}$ delivers only 98.42\% coverage even at maximum $EIRP_{\max}$, failing to satisfy the coverage constraint in \eqref{cov_c}.
This comparison highlights the necessity of the proposed EMF-aware BS deployment workflow. The empirical channel model characterizes the statistical dependence of path loss on distance without explicitly accounting for urban geometry. In the considered scenario, this leads to underestimated EMF exposure and a substantially different optimized deployment configuration. These discrepancies underscore the importance of incorporating deterministic channel modeling into BS deployment and power optimization, providing a more realistic basis for informed and reliable decisions in EMF-aware network planning and deployment.

\section{Conclusion}
\label{sec:conc}
In this paper, a practical workflow that determines the optimal BS deployment and radiated power is proposed for addressing the coverage-guaranteed \ac{EMF} exposure minimization problem in a 3D urban scenario. First, the novel least-time \ac{SBR} \ac{RL} algorithm is introduced for \ac{EMF} exposure calculation, with diffraction modeling reducing the \ac{RMSE} against real-world measurements from \SI{6.63}{dB} to \SI{1.29}{dB}, underscoring the importance of diffraction in our scenario and demonstrating the accuracy of the algorithm. Next, the \ac{AGR} algorithm substantially decreases the computational complexity of estimating \ac{EMF} exposure over the entire target area, requiring only 4.0\% of the simulation while achieving an accuracy of \SI{2.74}{dB} compared to the finest grid. Moreover, through system-level abstractions of the practical network behavior, including the actual transmit power, intercell interference and \ac{CSI} imperfection and the \ac{UE} mobility model, the aforementioned problem is formulated. Finally, after the convexification of the BS deployment feasible region, the problem is solved by the \ac{NM}-based method, which converges after 35 iterations. Simulation comparison with an empirical baseline in the considered urban scenario demonstrates that the proposed workflow mitigates the underestimation of EMF exposure and yields BS deployment configurations better aligned with real-world conditions, demonstrating its potential to offer valuable guidance for future EMF-aware BS deployment and urban planning.

The accuracy of EMF exposure estimation could be further improved by incorporating LiDAR-based 3D reconstruction to represent building surfaces and edges with greater geometric fidelity. To further reduce the cost of repeated channel simulations, learning-based surrogate models trained on deterministic channel results could also be explored and facilitate the use of a broader range of BS deployment optimization algorithms building on the proposed workflow.

\bibliography{mybibliography.bib} 
\bibliographystyle{ieeetr} 

\end{document}

%% file: acronyms.tex
\DeclareAcronym{3G}{short = 3G , long = third generation}
\DeclareAcronym{3GPP}{short = 3GPP , long = Third Generation Partnership Project}
\DeclareAcronym{4G}{short = 4G , long = fourth generation}
\DeclareAcronym{5G}{short = 5G , long = fifth generation}
\DeclareAcronym{6G}{short = 6G , long = sixth generation}

\DeclareAcronym{AoA}{short = AoA , long = angle of arrival}
\DeclareAcronym{AoD}{short = AoD , long = angle of departure}
\DeclareAcronym{AS}{short = AS , long = acceleration structure}
\DeclareAcronym{AGR}{short = AGR , long = adaptive grid refinement}

\DeclareAcronym{BBU}{short = BBU ,  long = baseband unit}
\DeclareAcronym{BS}{short = BS ,  long = base station}

\DeclareAcronym{CN}{short = CN , long = core network}
\DeclareAcronym{CPU}{short = CPU , long = central processing unit}
\DeclareAcronym{CH}{short = CH , long = closest hit}
\DeclareAcronym{CSI}{short = CSI , long = channel state information}
\DeclareAcronym{CDF}{short = CDF , long = cumulative distribution function}
\DeclareAcronym{CI}{short = CI , long = close-in}

\DeclareAcronym{EIRP}{short = EIRP , long = effective isotropic radiated power}
\DeclareAcronym{eMBB}{short = eMBB , long = enhanced mobile broadband}
\DeclareAcronym{EMF}{short = EMF , long = electromagnetic field}
\DeclareAcronym{EM}{short = EM , long = electromagnetic}
\DeclareAcronym{ETSI}{short = ETSI , long = European Telecommunications Standards Institute}
\DeclareAcronym{E-field}{short = E-field , long = electric field}
\DeclareAcronym{EI}{short = EI , long = exposure index}

\DeclareAcronym{FCC}{short = FCC , long = Federal Communications Commission}
\DeclareAcronym{FDD}{short = FDD , long = frequency division duplex}
\DeclareAcronym{FR1}{short = FR1 , long = frequency range 1}
\DeclareAcronym{FR2}{short = FR2 , long = frequency range 2}
\DeclareAcronym{FSPL}{short = FSPL , long = free-space path loss}

\DeclareAcronym{GO}{short = GO , long = geometrical optics}
\DeclareAcronym{GPU}{short = GPU , long = graphics processor unit}
\DeclareAcronym{GAS}{short = GAS , long = geometry acceleration structure}
\DeclareAcronym{GA}{short = GA , long = genetic algorithm}

\DeclareAcronym{HDII}{short = HDII , long = half-distance insersion and interpolation}
\DeclareAcronym{ICES}{short = ICES , long = International Committee on Electromagnetic Safety}
\DeclareAcronym{ICNIRP}{short = ICNIRP , long = International Commission on Non-Ionizing Radiation Protection}
\DeclareAcronym{IEC}{short = IEC , long = International Electrotechnical Commission}
\DeclareAcronym{IEEE}{short = IEEE , long = Institute of Electrical and Electronics Engineers}
\DeclareAcronym{IMT}{short = IMT , long = International Mobile Telecommunications}
\DeclareAcronym{IoT}{short = IoT , long = Internet of Things}
\DeclareAcronym{ITU}{short = ITU , long = International Telecommunication Union}
\DeclareAcronym{IHE}{short = IHE , long = Institute of Radio Frequency Engineering and Electronics}
\DeclareAcronym{IS}{short = IS , long = intersection}

\DeclareAcronym{KIT}{short = KIT , long = Karlsruhe Institute of Technology}

\DeclareAcronym{LoS}{short = LoS , long = line-of-sight}
\DeclareAcronym{LTE}{short = LTE , long = long term evolution}

\DeclareAcronym{mMIMO}{short = mMIMO , long = massive multiple input multiple output}
\DeclareAcronym{mmWave}{short = mmWave , long = millimeter wave}
\DeclareAcronym{mRRH}{short = mRRH , long = micro-RRH}
\DeclareAcronym{MSS}{short = MSS , long = mobile satellite service}
\DeclareAcronym{MIMO}{short = MIMO , long = multiple input multiple output}
\DeclareAcronym{MADS}{short = MADS , long = mesh adaptive direct search}
\DeclareAcronym{MRT}{short = MRT , long =  maximum ratio transmission}
\DeclareAcronym{MC}{short = MC , long =  Monte Carlo}
\DeclareAcronym{MMSE}{short = MMSE , long =  minimum mean square error}

\DeclareAcronym{NCRP}{short = NCRP , long = National Council on Radiation Protection and Measurements}
\DeclareAcronym{NG-RAN}{short = NG-RAN , long = next generation radio access network}
\DeclareAcronym{NR}{short = NR , long = new radio}
\DeclareAcronym{NSA}{short = NSA , long = non-stand-alone}
\DeclareAcronym{NTN}{short = NTN , long = non-terrestrial network}
\DeclareAcronym{NM}{short = NM , long = Nelder-Mead}
\DeclareAcronym{NLoS}{short = NLoS , long = non line-of-sight}

\DeclareAcronym{OSM}{short = OSM , long = OpenStreetMap}

\DeclareAcronym{PBCH}{short = PBCH , long = physical broadcast channel}
\DeclareAcronym{PCI}{short = PCI , long = physical cell ID}
\DeclareAcronym{PDSCH}{short = PDSCH , long = physical downlink shared channel}
\DeclareAcronym{pRRH}{short = pRRH , long = pico-RRH}
\DeclareAcronym{PTX}{short = PTX , long = parallel thread execution}
\DeclareAcronym{PDP}{short = PDP , long = power delay profile}
\DeclareAcronym{PLE}{short = PLE , long = path loss exponent}
\DeclareAcronym{PRF}{short = PRF , long =  power reduction factor}
\DeclareAcronym{PPP}{short = PPP , long =  Poisson Point Process}

\DeclareAcronym{QoE}{short = QoE , long =  quality of experience }
\DeclareAcronym{QoS}{short = QoS , long =  quality of service }

\DeclareAcronym{RAN}{short = RAN , long = radio access network}
\DeclareAcronym{RAT}{short = RAT , long = radio access technology}
\DeclareAcronym{RF}{short = RF , long = radio frequency}
\DeclareAcronym{RRH}{short = RRH , long = remote radio heads}
\DeclareAcronym{RSS}{short = RSS , long = root sum square}
\DeclareAcronym{RL}{short = RL , long = ray-launching}
\DeclareAcronym{RT}{short = RT , long = ray-tracing}
\DeclareAcronym{Rx}{short = Rx , long = receiver}
\DeclareAcronym{RMSE}{short = RMSE , long = root mean squared error}
\DeclareAcronym{RSSI}{short = RSSI , long = received signal strength indicator}
\DeclareAcronym{RSRP}{short = RSRP , long = reference signal received power}
\DeclareAcronym{RMS}{short = RMS , long = root mean square}
\DeclareAcronym{RIS}{short = RIS , long = reconfigurable intelligent surface}

\DeclareAcronym{SA}{short = SA , long = spectrum analyzer}
\DeclareAcronym{SAN}{short = SAN ,  long = satellite access node}
\DeclareAcronym{SAR}{short = SAR , long = specific absorption rate}
\DeclareAcronym{SSB}{short = SSB , long = synchronization signal block}
\DeclareAcronym{SBR}{short = SBR , long = shoot and bouncing ray}
\DeclareAcronym{SBT}{short = SBT , long = shader binding table}
\DeclareAcronym{std}{short = std , long = standard deviation}
\DeclareAcronym{SIR}{short = SIR , long = signal-to-interference ratio}
\DeclareAcronym{SNR}{short = SNR , long = signal-to-noise ratio}
\DeclareAcronym{SINR}{short = SINR , long = signal-to-interference-plus-noise ratio}
\DeclareAcronym{SG}{short = SG , long = stochastic-geometry}
\DeclareAcronym{SS-RSRP}{short = SS-RSRP , long = Synchronization Signals - Reference Signal Received Power}

\DeclareAcronym{TDD}{short = TDD , long = time division duplex}
\DeclareAcronym{THz}{short = THz , long = terahertz}
\DeclareAcronym{TN}{short = TN , long = terrestrial network}
\DeclareAcronym{TUK}{short = TUK, long = Technische Universit\"at Kaiserslautern}
\DeclareAcronym{Tx}{short = Tx , long = transmitter}

\DeclareAcronym{UAS}{short = UAS , long = unmanned aerial system }
\DeclareAcronym{UE}{short = UE , long = user equipment}
\DeclareAcronym{UTD}{short = UTD , long = uniform theory of diffraction}
\DeclareAcronym{UAV}{short = UAV , long = unmanned aerial vehicle}

\DeclareAcronym{VSAT}{short = VSAT , long = very small aperture terminal }
\DeclareAcronym{VPL}{short = VPL , long = vertical-plane-launch}

\DeclareAcronym{WHO}{short = WHO , long = World Health Organization}

\DeclareAcronym{ZF}{short = ZF , long = zero-forcing}